\documentclass[12pt,english,floatfix,superscriptaddress,aps,prd,preprint,nofootinbib]{revtex4}
\usepackage{amsmath}
\usepackage{amssymb}
\usepackage{amsbsy}
\usepackage[a4paper, margin=1.6cm]{geometry}
\usepackage{amsfonts}
\usepackage{amsopn}
\usepackage{amstext}
\usepackage{graphicx}
\usepackage{amssymb}
\usepackage{amsfonts}
\usepackage{amsmath}
\usepackage{graphicx}
\usepackage[english]{babel}
\usepackage{color}
\usepackage{xcolor}
\usepackage{slashed}
\usepackage{esint}
\usepackage[dvips]{epsfig}
\usepackage[dvips]{graphicx}
\usepackage{float}
\usepackage{units}
\usepackage{textcomp}
\usepackage{placeins}
\usepackage{hyperref}             
\hypersetup{
    colorlinks=true,              
    breaklinks=true,              
    citecolor=blue,               
    linkcolor=[rgb]{0,0.5,0.9},   
    urlcolor=red,                 
    filecolor=green               
}

\usepackage{hyperref}
\usepackage{slashed}

\newcommand{\ie}{\begin{equation}}
\newcommand{\fe}{\end{equation}}
 \newcommand{\bq}{\begin{equation}}
 \newcommand{\eq}{\end{equation}}
 \newcommand{\bqn}{\begin{eqnarray}}
 \newcommand{\eqn}{\end{eqnarray}}

\begin{document}

\title{Non-metricity effects on electron scattering in bumblebee gravity}



\author{A. A. Ara\'{u}jo Filho}
\email{dilto@fisica.ufc.br}
\affiliation{Departamento de Física, Universidade Federal da Paraíba, Caixa Postal 5008, 58051--970, João Pessoa, Paraíba,  Brazil.}
\affiliation{Departamento de Física, Universidade Federal de Campina Grande Caixa Postal 10071, 58429-900 Campina Grande, Paraíba, Brazil.}
\affiliation{Center for Theoretical Physics, Khazar University, 41 Mehseti Street, Baku, AZ-1096, Azerbaijan.}

\date{\today}

\begin{abstract}

We investigate \textit{non--metricity} effects on electron scattering in \textit{metric--affine} bumblebee gravity, where spontaneous Lorentz symmetry breaking is induced by a vector field acquiring a nonzero vacuum expectation value. Treating the affine connection as an independent variable and integrating it out leads to an effective description in which \textit{non--metricity} modifies the dispersion relation of the bumblebee modes. From the full momentum--space propagator, we determine the pole structure that governs the interaction and construct the corresponding static Green function and interparticle potential. For a purely timelike background, the dispersion relation remains isotropic and produces a Coulomb potential with a uniformly rescaled effective coupling; consequently, the scattering amplitude preserves the Rutherford angular dependence, with the Lorentz--violating parameter entering only as an overall multiplicative factor. In contrast, a spacelike background induces anisotropy in the dispersion relation, leading to an orientation--dependent potential characterized by a quadrupolar modulation. This anisotropic structure propagates to the differential and integrated cross sections, introducing directional dependence while preserving the long--range character of the interaction. Finally, we consider phenomenological constraints from atomic physics. Hydrogen spectroscopy constrains the isotropic sector associated with the timelike configuration, whereas searches for anisotropies provide stronger limits on the quadrupolar contribution governed by $\xi b^{2}$.

\end{abstract}

\maketitle
\tableofcontents


\section{Introduction}

Over the past decades, considerable effort has been devoted to investigating possible violations of Lorentz symmetry. Early indications emerged within string–based constructions, where extended objects such as D–branes can induce spontaneous breaking of spacetime symmetries in the higher–dimensional bulk \cite{Kostelecky:1989jp,Kostelecky:1989jw,Kostelecky:1991ak,Kostelecky:1994rn,Gliozzi:2011hj}. Related possibilities were later discussed in alternative quantum–gravity scenarios, including loop quantum gravity, spacetime foam models, and frameworks with modified dispersion relations \cite{Gambini:1998it,Calcagni:2016zqv,Bojowald:2004bb,Alfaro:2004aa,Amelino-Camelia:2001dbf}.

Since direct experimental access to Planck–scale dynamics remains unattainable, a pragmatic strategy is to encode potential high–energy remnants in effective field theories that parametrize deviations from local Lorentz invariance. In this context, the Standard–Model Extension (SME) \cite{Colladay:1996iz,Colladay:1998fq} provides a huge catalogue of Lorentz-- and CPT--violating operators constructed from conventional matter and gauge fields, later extended to include gravity within a low–energy general--relativistic description \cite{Kostelecky:2003fs,Liu:2025bpp}. These operators are typically interpreted as low–energy manifestations of an underlying theory in which tensor fields acquire nonzero vacuum expectation values, thereby selecting preferred spacetime directions.

When gravity is present, fixing such vacuum configurations to constant backgrounds generally restricts the admissible geometries, as illustrated in Einstein–aether models \cite{Jacobson:2000xp} and in Chern–Simons–modified gravity \cite{Jackiw:2003pm}. Promoting the symmetry–breaking fields to dynamical degrees of freedom avoids these consistency issues: once their equations of motion are satisfied, general covariance is preserved and the background structure arises dynamically rather than by prescription \cite{Jacobson:2000xp}.

Within the gravitational sector of the SME, several constructions have been proposed, but the so–called bumblebee scenario stands out from an effective–field–theory viewpoint \cite{Kostelecky:2003fs,Seifert:2009gi,Casana:2017jkc,Shi:2025rfq,Shi:2025plr,Ovgun:2018xys,AraujoFilho:2024iox,Amarilo:2023wpn,Maluf:2020kgf,Maluf:2014dpa}. Its action contains only curvature terms linear in the Ricci tensor, which dominate in the low–energy regime. The setup introduces a vector field subject to a symmetry–breaking potential; once it develops a nonzero vacuum expectation value, Lorentz invariance is spontaneously violated. This field couples non--minimally to gravity through contractions with the Ricci tensor. Extensions including higher–curvature invariants have also been examined, especially in contexts where strong–field effects may become relevant.

Most investigations of Lorentz--violating gravity adopt the purely metric formulation, assuming from the outset that the metric tensor fully characterizes spacetime geometry. An alternative viewpoint treats the metric and the affine connection as independent variables. In this \textit{metric--affine} (Palatini) framework, compatibility between connection and metric is not imposed {\it a priori}. 

Early analyses of Lorentz symmetry breaking in non–Riemannian settings explored backgrounds with torsion \cite{Kostelecky:2003fs} and \textit{non--metricity} \cite{Foster:2016uui}, as well as more general constructions such as Riemann–Finsler geometries \cite{Kostelecky:2011qz}. Building on these ideas, a \textit{metric--affine} realization of the bumblebee model was developed in \cite{Delhom:2019wcms}. Also, the weak--field stability of the resulting effective theory, including couplings to scalar and Dirac matter, was also established \cite{Delhom:2020gfv}. At the classical level, that formulation can be embedded in the broader class of Ricci--based gravities \cite{BeltranJimenez:2017doy,Afonso:2018bpv,Afonso:2018mxn,Delhom:2019zrb,BeltranJimenez:2019acz}, where matter fields couple non--minimally to the Ricci tensor.

In addition, \textit{non--metricity} has recently acquired a prominent role in black hole physics. It has been employed not only to generate new classes of black hole geometries, but also to investigate their thermodynamic, optical, and perturbative properties in a variety of contexts \cite{Jha:2023vhn,Lambiase:2023zeo,Shi:2025ywa,Heidari:2024bvd,Filho:2024isd,Ovgun:2025mil}.

Motivated by this framework, we investigate how \textit{non--metricity} affects electron scattering. The affine connection is treated as an independent variable and eliminated at the level of the field equations, yielding an effective description in which the vector vacuum expectation value modifies the dispersion relation of the propagating modes. The momentum–space propagator then determines the pole structure governing the interaction. Two vacuum configurations are analyzed. When the background vector is purely timelike, the dispersion relation remains isotropic; the Coulomb potential preserves its $1/r$ profile, with the Lorentz--violating parameter $\xi b^{2}$ entering only through a global rescaling of the effective coupling. The scattering amplitude therefore maintains the standard Rutherford angular dependence. In contrast, a spacelike vacuum expectation value induces anisotropy in the dispersion relation, producing a quadrupolar correction to the potential and transferring directional dependence to both differential and integrated cross sections, while the characteristic forward enhancement of long--range interactions is retained. At high energies, Mott and eikonal analyses confirm the persistence of the forward $1/\gamma^{2}$ behavior, with \textit{non--metricity} acting as a deformation of the coupling strength. Constraints from atomic spectroscopy and searches for anisotropic effects are also discussed, providing bounds on $\xi b^{2}$.


\section{Bumblebee gravity in the \textit{metric--affine} formalism}\label{sec2}

We begin by presenting the bumblebee action in the metric--affine formulation on a curved spacetime background \cite{Delhom:2019wcms,Delhom:2020gfv}
\begin{eqnarray}
	\nonumber S_{B}&=&\int \mathrm{d}^4 x\,\sqrt{-g}\Big[\frac{1}{2\kappa^2}\Big(R(\Gamma)+\xi \mathrm{B}^{\alpha} \mathrm{B}^{ b} R_{\alpha b}(\Gamma)\Big)-\frac{1}{4}\mathrm{B}^{\mu\nu}\mathrm{B}_{\mu\nu}-V(\mathrm{B}^{\mu}\mathrm{B}_{\mu}\mp b^2)\Big] +\\
	&+& \int \mathrm{d}^4 x\,\sqrt{-g}\mathcal{L}_{M}(g_{\mu\nu},\psi).
	\label{bumblebee}
\end{eqnarray} 
In the action, $R(\Gamma)$ and $R_{\mu\nu}(\Gamma)$ are constructed from an independent affine connection $\Gamma$, while $\mathcal{L}_{M}$ represents the matter sector. The vector field $\mathrm{B}_{\mu}$ enters through its field strength $\mathrm{B}_{\mu\nu}=2\partial_{[\mu}\mathrm{B}_{\nu]}$ and a self–interaction potential $V$. The gravitational coupling is fixed by $\kappa^{2}=8\pi G$, and $\xi$ denotes a small parameter associated with the inverse square of a high–energy scale.

The formulation follows the \textit{metric--affine }prescription, so the connection is not constrained {\it a priori} by the metric. The potential is chosen such that $\mathrm{B}_{\mu}$ develops a vacuum expectation value $b_{\mu}$ with $b^{2}>0$, thereby inducing spontaneous Lorentz symmetry breaking. As a result, low--energy observables can inherit effects from this preferred background configuration.

Only the symmetric part of the Ricci tensor contributes dynamically, ensuring projective invariance and preventing additional ghost–like modes from propagating in the \textit{affine} sector \cite{Aoki:2019rvi}. In the matter sector, fields are minimally coupled to the metric and do not interact explicitly with the connection. This simplifies the connection field equation, including its torsional component. For fermions, an axial torsion coupling may be incorporated; it modifies the hypermomentum but leaves the symmetric connection sector, which is central to our discussion, essentially unaffected.

By independently varying the action (\ref{bumblebee}) with respect to the metric, the \textit{affine} connection, and the bumblebee vector field, and performing the necessary manipulations, we arrive at the corresponding set of field equations, which can be written as follows:
\begin{eqnarray}
	\kappa^2  T_{\mu\nu} &=&G_{(\mu\nu)}(\Gamma)-\frac{\xi}{2}g_{\mu\nu}\bigg( \mathrm{B}^{\alpha}\mathrm{B}^{ b}R_{\alpha b}(\Gamma)\bigg)
	+2\xi\bigg(\mathrm{B}_{(\mu}R_{\nu) b}(\Gamma)\bigg)\mathrm{B}^{ b}, \label{Riccieq}\\
	0&=&\nabla_{\lambda}^{(\Gamma)}\bigg[\sqrt{-g}g^{\mu\alpha}\bigg(\delta^{\nu}_{\alpha}+\xi \mathrm{B}^{\nu}\mathrm{B}_{\alpha}\bigg)\bigg],\label{connectioneq}\\
	\nabla_{\mu}^{(g)}\mathrm{B}^{\mu\nu}&=&-\frac{\xi}{\kappa^2}g^{\nu\alpha}B^{ b}R_{\alpha b}(\Gamma)+2 V^{\prime}\mathrm{B}^{\nu},\label{bumblebeeeq}
	\label{PDE}
\end{eqnarray}
with the total stress--energy tensor being decomposed as $T_{\mu\nu}=T_{\mu\nu}^{M}+T_{\mu\nu}^{\mathrm{B}}$, separating the standard matter contribution from the sector associated with the bumblebee field. The term $T_{\mu\nu}^{\mathrm{B}}$ arises exclusively from the kinetic and potential structure of $\mathrm{B}_{\mu}$, since this field does not interact directly with the curvature tensor. They are written as 
\begin{eqnarray}
	T_{\mu\nu}^{M}&=&-\frac{2}{\sqrt{-g}}\frac{\delta(\sqrt{-g}\mathcal{L}_{M})}{\delta g^{\mu\nu}},\\
	T_{\mu\nu}^{\mathrm{B}}&=& \mathrm{B}_{\mu\sigma}\mathrm{B}_{\nu}^{\ \sigma}-\frac{1}{4}g_{\mu\nu}\mathrm{B}^{\alpha}_{\ \sigma}\mathrm{B}^{\sigma}_{\ \alpha}-V g_{\mu\nu}+2V^{\prime}\mathrm{B}_{\mu}\mathrm{B}_{\nu}.
\end{eqnarray}
An important feature of the connection field equation is that it contains no derivatives of the connection itself. Consequently, the \textit{affine} connection does not represent an independent propagating mode but instead plays the role of an auxiliary variable. Once Eq.~(\ref{connectioneq}) is solved, the connection can be eliminated in favor of other fields and substituted back into the action.

Solving this algebraic relation shows that the affine structure coincides with the Levi--Civita connection of an effective metric $h^{\mu\nu}$, which is introduced as follows:
\begin{equation}
	h^{\mu\nu}=\frac{1}{\sqrt{1+\xi X}}(g^{\mu\nu}+\xi B^{\mu}B^{\nu}).\label{asfsdfs}
\end{equation}  
Here, the scalar quantity $X$ is defined by the contraction $X \equiv g^{\mu\nu} \mathrm{B}_{\mu} \mathrm{B}_{\nu}$. After expressing the \textit{affine} connection in terms of the effective metric and implementing suitable redefinitions of the fields, the original action (\ref{bumblebee}) can be rewritten in a form analogous to the Einstein--Hilbert action, namely:
\begin{eqnarray}
\label{essinbee}
	\tilde{\mathcal{S}}_{BEF}&=&\int \mathrm{d}^{4} x \sqrt{-h}\frac{1}{2 \kappa^{2}} R(h)+\overline{\mathcal{S}}_{m}\left(h_{\mu \nu}, \mathrm{B}_{\mu}, \psi\right)\;.
\end{eqnarray} 
The action in Eq.~(\ref{essinbee}) assumes the form of an Einstein–Hilbert theory built from the effective metric $h^{\mu\nu}$, while the matter sector is reshaped into a modified functional $\bar{\mathcal{S}}_{\rm m}$. This new matter contribution incorporates non–linear couplings between the bumblebee field and the original matter fields contained in $\mathcal{S}_{\rm m}$.

In this representation, the causal structure relevant for tensor perturbations is determined by $h_{\mu\nu}$. Accordingly, gravitational waves follow the null cones of this effective metric. Since the physical metric $g_{\mu\nu}$ remains minimally coupled to standard model fields, strong background configurations of the bumblebee field may lead to apparent subluminal or superluminal behavior of the metric perturbations associated with $g_{\mu\nu}$. Assessing whether present gravitational–wave observations can place quantitative bounds on such effects requires a dedicated analysis, which we postpone to future work.

Inspection of the connection field equation shows that a nonvanishing vacuum configuration of the bumblebee field generically induces a background value for the \textit{non--metricity} tensor, defined as $Q_{\mu}{}^{\alpha\beta} \equiv \nabla_{\mu} \,g^{\alpha\beta}$. In this way, spontaneous Lorentz symmetry breaking is accompanied by a dynamically generated non–metric geometric structure, in line with the mechanism proposed in \cite{Foster:2016uui}. Turning to the dynamics of the vector field itself, its equation of motion may be obtained either by inserting (\ref{asfsdfs}) into (\ref{bumblebeeeq}) or by varying directly the Einstein–frame action (\ref{essinbee}). In both cases, the resulting expression takes the form of an effective Proca–type equation for the bumblebee field
\begin{equation}
	\nabla_{\nu}^{(g)} \mathrm{B}^{\nu\mu}=\mathcal{M}^{\mu}_{\,\,\nu}\mathrm{B}^{\nu},
	\label{Proca}
\end{equation}
with the symbol $\nabla_{\nu}^{(g)}$ refers to the covariant derivative constructed from the Levi--Civita connection associated with the metric $g_{\mu\nu}$. The vector field acquires an effective mass structure encoded in the tensor $\mathcal{M}^{\mu}{}_{\nu}$, whose explicit form reads as follows:
\ie	
\mathcal{M}^{\mu}{}_{\nu} =\left(\frac{\xi T}{2+3\xi X}+\frac{\xi^2 \mathrm{B}^{\alpha}\mathrm{B}^{ b}T_{\alpha b}}{(1+\xi X)(2+3\xi X)}+2V^{\prime}\right)\delta^{\mu}_{\,\,\nu}- \frac{\xi}{(1+\xi X)}T^{\mu\alpha}g_{\nu\alpha} \ .
\fe
The tensor $\mathcal{M}^{\mu}{}_{\nu}$ contains the contribution $2V'$ that acts as a mass term, together with additional self--interactions and couplings to the remaining matter fields through the stress--energy tensor. Its structure is not necessarily positive definite, since $\det(\mathcal{M}^{\mu}{}_{\nu})$ can change sign, and therefore stability of the corresponding field equation is not automatically ensured.

The field equations obtained here differ markedly from those arising when the connection is assumed to be Levi–Civita from the beginning. In that case, enforcing compatibility between the metric and the connection leads, after variation, to additional derivative terms generated by integrations by parts. These terms introduce derivatives of the bumblebee field into the gravitational sector and, via contractions such as $\mathrm{B}_{\alpha}R^{\alpha\nu}$, modify the structure of the vector equation, producing a dynamics unlike Eq.~(\ref{Proca}).

By contrast, in the present construction the Einstein–frame form (\ref{essinbee}) makes explicit that the auxiliary metric $h_{\mu\nu}$ couples to the total energy–momentum content, with the weak–field limit characterized by $h_{\mu\nu}\simeq\eta_{\mu\nu}$. Since $g_{\mu\nu}$ depends on $X$ and on the disformal combination $\mathrm{B}_{\mu}\mathrm{B}_{\nu}$ (see Eq. (\ref{asfsdfs})), additional couplings between the bumblebee and matter fields emerge that are absent in the alternative formulation. The two setups therefore define genuinely distinct theories.


\section{Pole structure of the propagator}\label{sec3}

The analysis is performed in the weak–field regime, where the spacetime geometry is close to Minkowski. In this limit one has $h_{\mu\nu}\simeq\eta_{\mu\nu}$. Keeping only the leading terms in the Lorentz-violating parameter and discarding $\mathcal{O}(\xi^2)$ contributions, the metric may be approximated as $g_{\mu\nu}\simeq\eta_{\mu\nu}+\xi\!\left(B_\mu B_\nu-\frac{1}{2}X\,\eta_{\mu\nu}\right)$.
To proceed, a specific choice for the self-interaction potential of the bumblebee field must be introduced. Here we adopt the standard bumblebee potential, given by
\begin{equation}
	V(\mathrm{B}^\mu \mathrm{B}_\mu\mp b^2)=\frac{\lambda}{4}\big(\mathrm{B}^\mu \mathrm{B}_\mu\mp b^2\big)^2,
\end{equation}
Here $\lambda$ represents a small positive coupling parameter. The condition $b^{2}>0$ is assumed, while the symbol $\mp$ distinguishes between spacelike and timelike realizations of the bumblebee vacuum expectation value. After rewriting $g_{\mu\nu}$ in terms of $h_{\mu\nu}$ and $B_\mu$, additional effective self--interaction terms for the bumblebee field appear. In this formulation, the bumblebee {\color{red}Lagragian} in the Einstein frame becomes
\begin{eqnarray}
	\nonumber{\cal L}_{BEF}&=&-\frac{1}{4}\mathrm{B}_{\mu\nu}\mathrm{B}^{\mu\nu}+\frac{M^2}{2}\mathrm{B}^2-\frac{\Lambda}{4}(B^2)^2+\\ &+&\frac{\xi}{2}\Big[\mathrm{B}^{\mu\nu}\mathrm{B}^\alpha{}_{\nu}\mathrm{B}_\mu \mathrm{B}_\alpha-\frac{1}{4}\mathrm{B}_{\mu\nu}\mathrm{B}^{\mu\nu}\mathrm{B}^2-\frac{3}{4}\Lambda(\mathrm{B}^2)^3\Big]+
	\mathcal{O}(\xi^2).
	\label{BumbLagPert}
\end{eqnarray} 
The effective mass of the bumblebee field is defined as $M^2=\lambda b^2(\pm 1+\frac{\xi}{4}b^2)$, while the quartic coupling parameter is $\Lambda=\lambda(1\pm 2\xi b^2)$. From this point forward, all indices are manipulated with the Minkowski metric, implying $\mathrm{B}^2=\eta_{\mu\nu}\mathrm{B}^\mu\mathrm{B}^\nu$. This choice is consistent with the weak field treatment, where contributions of order $\mathcal{O}(\xi^2)$ are neglected.

Additional interaction terms appear as a consequence of the non--minimal coupling between the bumblebee field and the affine connection through the Ricci tensor. Such nonlinear structures are absent in the purely metric formulation, since in that case no direct coupling between the bumblebee sector and matter sources is present, as noted previously. By construction, cubic interaction vertices do not arise in this setup.

The spontaneous violation of Lorentz symmetry is assumed to originate from the self-interaction potential of the vector field, which is taken as $V=-\frac{M^2}{2}\mathrm{B}^2+\frac{\Lambda}{4}(\mathrm{B}^2)^2$. Higher--order operators induced by non--metricity carry additional powers of $\xi$ and are therefore suppressed within the present approximation. This type of potential has been examined at tree level in Ref.~\cite{Bluhm:2008yt}, where different classes of bumblebee potentials are systematically discussed.

Consider a background configuration in which the bumblebee field acquires a vacuum expectation value $\langle B^\mu\rangle=b^\mu$, satisfying $b^\mu b_\mu=\pm b^2$. Small fluctuations around this vacuum can be described by decomposing the field as $\mathrm{B}_\mu=b_\mu+\tilde{\mathrm{B}}_\mu$. Substituting this expansion into the Lagrangian in Eq.~(\ref{BumbLagPert}) and retaining the terms governing the perturbations yields the effective Lagrangian for the fluctuation field $\tilde{\mathrm{B}}_\mu$
\ie
\begin{split}
&\mathcal{L}_{BEF}^{\rm pert}=-\frac{1}{4}\tilde{\eta}_{\mu\alpha}\tilde{\mathrm{B}}^{\mu}_{\,\,\nu}\tilde{B}^{\alpha\nu}-\Lambda( b_\mu\tilde{\mathrm{B}}^\mu)^2-\frac{\Lambda}{4}(\tilde{\mathrm{B}}^2)^2 \\
&+\xi\Bigg[\frac{1}{2}\tilde{B}^{\mu\nu}\tilde{\mathrm{B}}^\alpha{}_\nu\tilde{\mathrm{B}}_\alpha\tilde{\mathrm{B}}_{\mu}-\frac{1}{4}\tilde{\mathrm{B}}^{\mu\nu}\tilde{\mathrm{B}}_{\mu\nu}( b_\rho\tilde{\mathrm{B}}^\rho)
+\tilde{\mathrm{B}}^{\mu\nu}\tilde{\mathrm{B}}^\alpha{}_\nu\tilde{\mathrm{B}}_\alpha b_{\mu}-\frac{1}{8}\tilde{\mathrm{B}}^{\mu\nu}\tilde{\mathrm{B}}_{\mu\nu}\tilde{\mathrm{B}}^2\Bigg]+\mathcal{O}(\xi^2,\lambda\xi),
\label{BumbLagVEV}
\end{split}
\fe
In this case, the kinetic term of the fluctuation field is fundamentally influenced by the background configuration. The perturbations therefore propagate in the presence of $b^\mu$, with their kinetic structure effectively coupled to a background dependent metric
\begin{eqnarray}
\tilde{\eta}_{\mu\nu}\equiv\eta_{\mu\nu}\left(1+\frac{\xi b^2}{2}\right)-2\xi  b_{\mu} b_{\nu}.
\end{eqnarray}
Because the vacuum configuration is nontrivial, the kinetic sector of the fluctuation field is modified. In particular, the Maxwell--type contribution acquires an overall factor, taking the form $-\frac{1}{4}(1+\frac{\xi b^2}{2})\,\tilde{\mathrm{B}}_{\mu\nu}\tilde{\mathrm{B}}^{\mu\nu}$, while an additional aether--like structure appears in the effective theory.

From this Lagrangian, the linearized field equations governing the vector perturbations follow as
\begin{equation}
\nonumber 0=\partial_{\mu}\tilde{\mathrm{B}}^{\mu\nu}\left(1+\frac{\xi b^2}{2}\right)-\xi  b_{\mu} b_{\alpha}\partial^{\mu}\tilde{\mathrm{B}}^{\alpha\nu}+\xi  b^{\nu} b_{\alpha}\partial_{\mu}\tilde{\mathrm{B}}^{\alpha\mu}-
2\Lambda b^{\nu} b_{\alpha}\tilde{\mathrm{B}}^{\alpha}.
\end{equation}
Equivalently, the same equations can be rewritten using the effective metric, in which case they take the form
$\tilde{\eta}_{\mu[\nu}\partial^{\nu}\tilde{\mathrm{B}}^{\mu}_{\,\,\alpha]}+M_{\alpha\mu}\tilde{\mathrm{B}}^{\mu}=0$.
Here $M_{\alpha\mu}=-2\Lambda b_\alpha b_\mu$ plays the role of an effective mass squared tensor. Taking the divergence of the field equation, which must vanish, yields a constraint that differs from the usual Maxwell condition $(\partial\!\cdot\!B)=0$. Instead, one obtains $(b\!\cdot\!\partial)(b\!\cdot\!\tilde B)=0$. This structure reflects the aether-like nature of the mass term, since the standard Proca contribution is absent. When this constraint is imposed, the free vector field action reduces to
\begin{eqnarray}
{\cal L}&=&-\frac{1}{4}\tilde{\mathrm{B}}_{\mu\nu}\tilde{\mathrm{B}}^{\mu\nu}\left(1+\frac{\xi b^2}{2}\right)-\frac{\xi}{2}\tilde{\mathrm{B}}_{\mu}[\Box b^{\mu} b^{\nu}+
( b\cdot\partial)^2\eta^{\mu\nu}]\tilde{\mathrm{B}}_{\nu}-\Lambda( b^{\alpha}\tilde{\mathrm{B}}_{\alpha})^2.
\end{eqnarray}

The parameter $\xi$ is taken to be small. Since it carries dimensions of inverse mass squared, it can be associated with the inverse square of a large energy scale. Under this assumption, only terms linear in $\xi$ are retained. At this order the relevant quantum corrections arise from single-vertex diagrams contributing to the two-point function, which implies that only quartic interaction vertices are required. Their evaluation requires the propagators in the presence of the background field, which take the form
\begin{eqnarray}
\langle\tilde{\mathrm{B}}^{\mu}(-k);\tilde{\mathrm{B}}^{\nu}(k)\rangle&=&
i\bigg[(-k^2\eta^{\mu\nu}+k^{\mu}k^{\nu})\Big(1+\frac{\xi b^2}{2}\Big)+
\xi(k^2 b^{\mu} b^{\nu}+( b\cdot k)^2\eta^{\mu\nu})-
2\Lambda b^{\mu} b^{\nu}\bigg]^{-1}.
\end{eqnarray}
The propagators follow from the explicit inversion of the corresponding operator matrices. These inverse matrices can be obtained directly and take the form
\begin{eqnarray}
(A\eta^{\alpha b}+Bk^{\alpha}k^{ b}+C  b^{\alpha} b^{ b})^{-1}&=&
\frac{1}{A}\eta_{\alpha b}-
\frac{(A+C b^2)B}{A\Delta}k_{\alpha}k_{ b}-\\
&-&\frac{(A+Bk^2)C}{A\Delta} b_{\alpha} b_{ b}+
\frac{BC}{A\Delta}( b\cdot k)( b_{\alpha}k_{ b}+ b_{ b}k_{\alpha}).
\end{eqnarray}
Here, $\Delta=(A+Bk^2)(A+Cb^2)-BC(b\!\cdot\!k)^2$. In the limit $\xi=0$, corresponding to the absence of non--metricity effects, this structure reduces to the expression reported in Ref.~\cite{Altschul:2005mu}. For the present setup the coefficients are given by $A=-k^2(1+\frac{\xi b^2}{2})+\xi(b\!\cdot\!k)^2$, $B=1+\frac{\xi b^2}{2}$, and $C=-2\Lambda+\xi k^2$. Using these definitions, the propagators can be written as follows
\ie
\begin{split}
& \langle\tilde{\mathrm{B}}_{\alpha}(-k);\tilde{B}_{ b}(k)\rangle=i\frac{1}{-k^2(1+\frac{\xi b^2}{2})+\xi( b\cdot k)^2}\Big\{\eta_{\alpha b}-
\Delta^{-1}\Big[\Big(-k^2\Big(1-\frac{1}{2}\xi b^2\Big)+\xi( b\cdot k)^2-\\
&-2\Lambda b^2\Big)\Big(1+\frac{\xi b^2}{2}\Big)k_{\alpha}k_{ b}+
\xi( b\cdot k)^2(-2\Lambda+\xi k^2) b_{\alpha} b_{ b}- 
\Big(1+\frac{\xi b^2}{2}\Big)
(-2\Lambda+\xi k^2)( b\cdot k)( b_{\alpha}k_{ b}+ b_{ b}k_{\alpha})
\Big]\Big\},
\end{split}
\fe
in which
\begin{eqnarray}
\Delta=-(-2\Lambda+\xi k^2)( b\cdot k)^2\Big(1+\frac{\xi b^2}{2}\Big)+
\xi( b\cdot k)^2\Big[-k^2\Big(1-\frac{1}{2}\xi b^2\Big)-2\Lambda b^2+\xi( b\cdot k)^2\Big].
\end{eqnarray}

From this point onward, we focus on the pole condition associated with the denominator $A(k)$ of the propagator,
\begin{equation}
A(k)=-k^2\left(1+\frac{\xi b_\mu b^\mu}{2}\right)+\xi (b\cdot k)^2=0,
\end{equation}
which we use as the effective dispersion relation for the propagating mode under consideration.

\subsection{Timelike configuration: $b^\mu=(b,0,0,0)$}

For a purely timelike background,
\begin{equation}
b^\mu=(b,0,0,0),
\qquad
b_\mu b^\mu=b^2,
\qquad
b\cdot k=b\omega,
\qquad
k^2=\omega^2-|\mathbf{k}|^2.
\end{equation}
Substituting into $A(k)=0$, we get
$-\left(1+\frac{\xi b^2}{2}\right)(\omega^2-|\mathbf{k}|^2)+\xi b^2\omega^2=0$,
which yields
$\omega^2\left(1-\frac{\xi b^2}{2}\right)
=
|\mathbf{k}|^2\left(1+\frac{\xi b^2}{2}\right)$.
Then,
$\omega^2=
\frac{1+\frac{\xi b^2}{2}}{1-\frac{\xi b^2}{2}}
\,|\mathbf{k}|^2$.
At first order in $\xi b^2$,
\begin{equation}
\omega^2\simeq (1+\xi b^2)|\mathbf{k}|^2.
\end{equation}
The timelike background preserves spatial isotropy, so the phase and group velocities are modified only by an overall direction-independent rescaling. Reality of the frequency requires $1-\xi b^2/2>0$, which is automatically satisfied in the perturbative regime $|\xi b^2|\ll1$.


\subsection{Spacelike configuration: $b^\mu=(0,\mathbf{b})$}

For a purely spacelike background, it is convenient to define the positive quantity
$|\mathbf{b}|^2\equiv \mathbf{b}\cdot\mathbf{b},
\,\,
b_\mu b^\mu=-|\mathbf{b}|^2$.
Then
$b^\mu=(0,\mathbf{b}),
\,\,
b\cdot k=-\mathbf{b}\cdot\mathbf{k},
\,\,
k^2=\omega^2-|\mathbf{k}|^2$.
Substituting into $A(k)=0$, we obtain
\begin{equation}
-\left(1-\frac{\xi |\mathbf{b}|^2}{2}\right)(\omega^2-|\mathbf{k}|^2)
+\xi(\mathbf{b}\cdot\mathbf{k})^2=0.
\end{equation}
Solving for $\omega^2$ gives
\begin{equation}
\omega^2=
|\mathbf{k}|^2+
\frac{\xi}{1-\frac{\xi |\mathbf{b}|^2}{2}}
(\mathbf{b}\cdot\mathbf{k})^2.
\end{equation}
To first order in $\xi |\mathbf{b}|^2$, we have $\omega^2\simeq
|\mathbf{k}|^2+\xi(\mathbf{b}\cdot\mathbf{k})^2$.
Also, writing
$(\mathbf{b}\cdot\mathbf{k})^2=|\mathbf{b}|^2\,|\mathbf{k}|^2\cos^2\vartheta$,
with $\vartheta$ the angle between $\mathbf{k}$ and $\mathbf{b}$, one may rewrite this as
\begin{equation}
\omega^2\simeq
|\mathbf{k}|^2\left(1+\xi |\mathbf{b}|^2\cos^2\vartheta\right).
\end{equation}
The spacelike background therefore induces an anisotropic dispersion relation, so propagation depends on the orientation of the wave vector relative to the preferred direction.


\section{The corresponding interparticle potentials}

\subsection{Timelike configuration}

\begin{figure}
    \centering
     \includegraphics[scale=0.6]{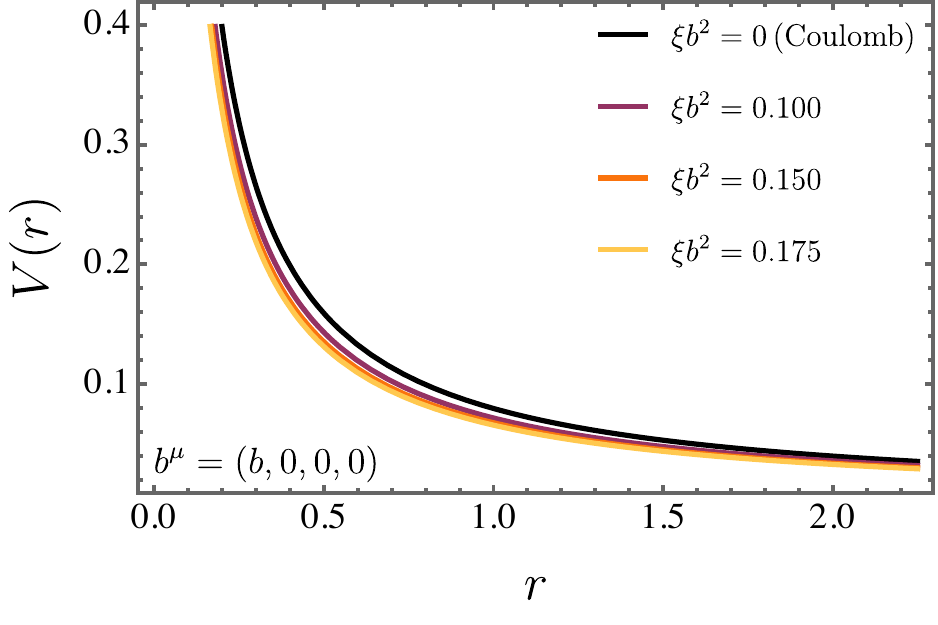}
    \caption{The interparticle potential $V(r)$ in the timelike configuration as a function of the radial distance $r$, for different values of the parameter $\xi b^{2}$.}
    \label{timelikev}
\end{figure}

To obtain the interaction potential in coordinate space, one performs the inverse Fourier transform of the propagator in momentum space. Assuming spherical symmetry, the angular integrations can be carried out explicitly, which reduces the three--dimensional integral to a one--dimensional radial integral over the momentum magnitude. The resulting expression reads
\ie
\begin{split}
\label{interpoott}
V(r) &= \int \frac{\mathrm{d}^{3}k}{(2\pi)^{3}}\, e^{i\mathbf{k}\cdot\mathbf{r}}\, G(k)\\
&= \frac{1}{(2\pi)^3}\int_{0}^{\infty} \mathrm{d}k\, k^2 \int_{0}^{\pi} \mathrm{d}\theta \sin\theta \int_{0}^{2\pi} \mathrm{d}\varphi\,
e^{ikr\cos\theta}\,G(k)= \frac{1}{2\pi^{2}r}\int_{0}^{\infty} \mathrm{d}k\, k \sin(kr)\,G(k).
\end{split}
\fe
For the timelike configuration, the static limit of the propagator is obtained by setting $\omega=0$ in the denominator
$A(k)=-k^2\left(1+\frac{\xi b^2}{2}\right)+\xi (b\cdot k)^2$.
Since $b\cdot k=b\omega$, we find
$A(\omega=0,\mathbf{k})=\left(1+\frac{\xi b^2}{2}\right)k^2$,
and therefore
\ie
\begin{split}
V(r)
&= \frac{1}{2\pi^{2}r}\int_{0}^{\infty} \mathrm{d}k\, k \sin(kr)\,
\frac{1}{\left(1+\frac{\xi b^{2}}{2}\right)k^{2}}= \frac{1}{4\pi r}\,\frac{1}{1+\frac{\xi b^{2}}{2}}
\approx \frac{1}{4\pi r}-\frac{\xi b^{2}}{8\pi r}.
\label{vtimeee}
\end{split}
\fe
The resulting interparticle potential preserves the standard long--range Coulombic behavior, scaling as $V(r)\propto 1/r$, and therefore does not introduce any Yukawa--type screening or additional radial structure. Instead, the \textit{metric--affine} bumblebee background produces a uniform multiplicative rescaling of the interaction strength through the factor $\left(1+\frac{\xi b^{2}}{2}\right)^{-1}$. This modification can be interpreted as an effective renormalization of the coupling constant, such that the Lorentz--violating parameter $\xi b^{2}$ shifts the overall magnitude of the interaction without altering its spatial profile. For small $\xi b^{2}$, the expansion
$V(r)\simeq \frac{1}{4\pi r}\left(1-\frac{\xi b^{2}}{2}\right)$
shows that the correction is linear in the Lorentz--violating background and amounts to a fractional change in the interaction strength. In Fig.~\ref{timelikev}, we display the interparticle potential $V(r)$ for the timelike configuration. The curves show that the magnitude of the potential decreases as the Lorentz--violating parameter increases, while its Coulombic radial profile is preserved.

\subsection{Spacelike configuration}

Proceeding in the same way as in the timelike case, we write
\ie
\begin{split}
\label{interpo_spacelike}
V(\mathbf r)
&= \int \frac{\mathrm d^{3}k}{(2\pi)^{3}}\,
e^{i\mathbf{k}\cdot\mathbf{r}}\,
G(\mathbf k)\,.
\end{split}
\fe
For the spacelike configuration, the dispersion relation obtained from the pole condition is
$\omega^{2}\simeq k^{2}\Big(1+\xi b^{2}\cos^{2}\vartheta\Big),
\,\,
a\equiv \xi b^{2}$,
where $b^{2}\equiv |\mathbf b|^{2}>0$. However, for full consistency with the propagator, the static Green kernel must be constructed from the complete denominator before imposing the on--shell approximation. From the pole structure, one has
$A(\omega,\mathbf k)
=
-\left(1-\frac{a}{2}\right)(\omega^{2}-k^{2})
+a(\mathbf k\cdot\hat{\mathbf b})^{2}$,
where $\hat{\mathbf b}$ denotes the unit vector along the spacelike background. In the static limit $\omega=0$, this becomes
$A(0,\mathbf k)
=
\left(1-\frac{a}{2}\right)k^{2}
+a(\mathbf k\cdot\hat{\mathbf b})^{2}$.
Therefore, the corresponding static Green kernel is
\begin{equation}
\label{Gk_aniso}
G(\mathbf k)
=
\frac{1}{
\left(1-\frac{a}{2}\right)k^{2}
+a(\mathbf k\cdot\hat{\mathbf b})^{2}
}.
\end{equation}

Since $
(\mathbf k\cdot\hat{\mathbf b})^{2}=k^{2}\cos^{2}\vartheta$,
the kernel may also be written as
\begin{equation}
G(\mathbf k)
=
\frac{1}{k^{2}\left(1-\frac{a}{2}+a\cos^{2}\vartheta\right)}
=
\frac{1}{1-\frac{a}{2}}\,
\frac{1}{k^{2}\left(1+\dfrac{a}{1-a/2}\cos^{2}\vartheta\right)}.
\end{equation}

Since $\hat{\mathbf b}$ selects a preferred spatial direction, the potential is no longer purely radial. It depends on the angle $\hat{\alpha}$ between $\mathbf r$ and $\hat{\mathbf b}$, $\cos\hat{\alpha} \equiv \hat{\mathbf r}\cdot\hat{\mathbf b},
\,\,
V(\mathbf r)=V(r,\hat{\alpha})$.

Choosing coordinates such that $\hat{\mathbf b}$ defines the $z$--axis, we have
$\cos\vartheta=\hat{\mathbf k}\cdot\hat{\mathbf b},
\,\,
\mathbf k\cdot\mathbf r = kr\cos\beta$,
with
$\cos\beta
=
\cos\vartheta\cos\hat{\alpha}+\sin\vartheta\sin\hat{\alpha}\cos\varphi$. Substituting \eqref{Gk_aniso} into \eqref{interpo_spacelike} and using spherical variables $(k,\vartheta,\varphi)$ defined with respect to $\hat{\mathbf b}$, we obtain
\begin{equation}
\label{V_before_phi}
V(r,\hat{\alpha})
=
\frac{1}{(2\pi)^3}\int_{0}^{\infty}\! \mathrm dk\,k^{2}
\int_{0}^{\pi}\!\mathrm d\vartheta\,\sin\vartheta
\int_{0}^{2\pi}\!\mathrm d\varphi\;
\frac{e^{ikr\cos\beta}}{k^{2}\left(1-\frac{a}{2}+a\cos^{2}\vartheta\right)}.
\end{equation}

The azimuthal integral can be performed exactly:
\begin{equation}
\label{phi_identity}
\int_{0}^{2\pi}\!\mathrm d\varphi\;e^{ikr\cos\beta}
=
2\pi\,e^{ikr\cos\hat{\alpha}\cos\vartheta}\,
J_{0}\!\big(kr\sin\hat{\alpha}\sin\vartheta\big),
\end{equation}
where $J_{0}(x)$ denotes the Bessel function of the first kind of order zero. Therefore,
\begin{equation}
\label{V_Bessel_form}
V(r,\hat{\alpha})
=
\frac{1}{(2\pi)^2}\int_{0}^{\infty}\! \mathrm dk
\int_{0}^{\pi}\!\mathrm d\vartheta\,\sin\vartheta\;
\frac{
e^{ikr\cos\hat{\alpha}\cos\vartheta}\,
J_{0}\!\big(kr\sin\hat{\alpha}\sin\vartheta\big)
}{
1-\frac{a}{2}+a\cos^{2}\vartheta
}.
\end{equation}

A closed--form expression follows by observing that, in the frame where $\hat{\mathbf b}=\hat{\mathbf z}$, $\left(1-\frac{a}{2}\right)k^{2}+a(\mathbf k\cdot\hat{\mathbf b})^{2}
=
\left(1-\frac{a}{2}\right)(k_x^{2}+k_y^{2})
+
\left(1+\frac{a}{2}\right)k_z^{2}$. Then, $V$ satisfies the anisotropic Poisson equation
\begin{equation}
\left[
\left(1-\frac{a}{2}\right)(\partial_x^{2}+\partial_y^{2})
+
\left(1+\frac{a}{2}\right)\partial_z^{2}
\right]V(\mathbf r)
=
-\,\delta^{(3)}(\mathbf r).
\end{equation}
Introducing the rescaled coordinates
$x=\sqrt{1-\frac{a}{2}}\,x',
\,\,
y=\sqrt{1-\frac{a}{2}}\,y',
\,\,
z=\sqrt{1+\frac{a}{2}}\,z'$, the operator becomes isotropic in $(x',y',z')$, and the Green function is
\begin{equation}
V(\mathbf r)
=
\frac{1}{4\pi\left(1-\frac{a}{2}\right)\sqrt{1+\frac{a}{2}}}
\frac{1}{
\sqrt{
\frac{x^{2}+y^{2}}{1-a/2}
+
\frac{z^{2}}{1+a/2}
}
}.
\end{equation}

Writing $z=r\cos\hat{\alpha}$, we obtain
\begin{equation}
\label{V_exact_spherical}
V(r,\hat{\alpha})
=
\frac{1}{4\pi r}\,
\frac{1}{
\sqrt{\left(1-\frac{a}{2}\right)\left(1+\frac{a}{2}-a\cos^{2}\hat{\alpha}\right)}
}.
\end{equation}
Equivalently,
\begin{equation}
V(r,\hat{\alpha})
=
\frac{1}{4\pi r}\,
\frac{1}{1-\frac{a}{2}}\,
\frac{1}{
\sqrt{
1+\dfrac{a}{1-a/2}\sin^{2}\hat{\alpha}
}
}.
\end{equation}

\begin{figure}
    \centering
     \includegraphics[scale=0.45]{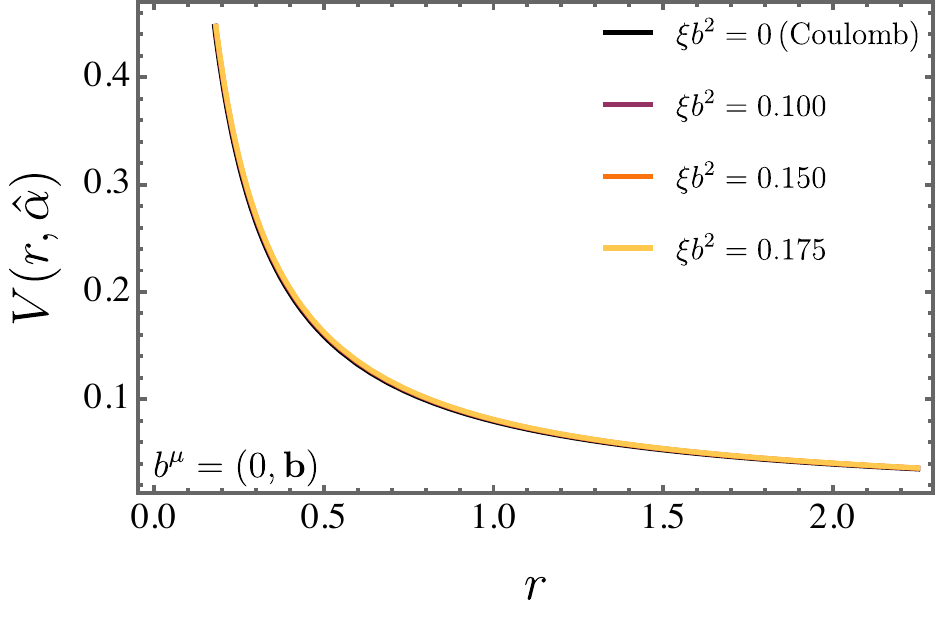}
     \includegraphics[scale=0.45]{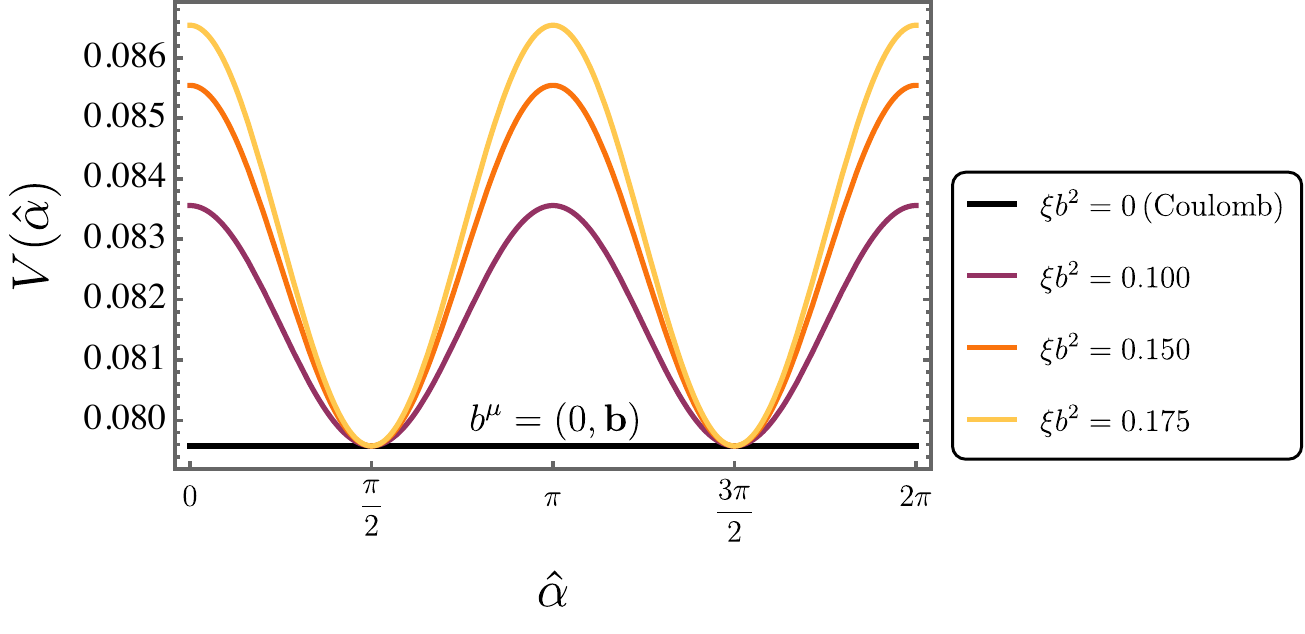}
    \caption{Radial and angular behavior of the modified potential $V(r,\hat{\alpha})$ in the spacelike configuration. The left panel shows the radial profile for $\hat{\alpha}=\pi/3$ and different values of $\xi b^{2}$, where the $1/r$ decay is preserved with a rescaled amplitude. The right panel presents the angular dependence at $r=1$.}
    \label{vspacelike}
\end{figure}

Expanding for small $|a|\ll1$, one finds
\begin{equation}
\label{V_small_a}
V(r,\hat{\alpha})
=
\frac{1}{4\pi r}
\left[
1+\frac{a}{2}\cos^{2}\hat{\alpha}
+O(a^{2})
\right]
=
\frac{1}{4\pi r}
\left[
1+\frac{a}{6}
+\frac{a}{3}P_{2}(\cos\hat{\alpha})
+O(a^{2})
\right],
\end{equation}
where $P_{2}(x)=\dfrac{1}{2}(3x^{2}-1)$ denotes the second Legendre polynomial. Restoring $a=\xi b^{2}$, the final result reads
\begin{equation}
V(r,\hat{\alpha})
=
\frac{1}{4\pi r}
\left[
1+\frac{\xi b^{2}}{6}
+\frac{\xi b^{2}}{3}P_{2}(\cos\hat{\alpha})
+O\!\big((\xi b^{2})^{2}\big)
\right].
\end{equation}
which are automatically satisfied in the perturbative regime $|\xi b^{2}|\ll1$. These features are illustrated in Fig.~\ref{vspacelike}. 
The potential preserves the Coulombic radial decay $1/(4\pi r)$, so the interaction remains long ranged and massless, but it acquires an anisotropic dependence controlled by the angle $\hat{\alpha}$ relative to the preferred direction selected by $b^\mu$. For $\xi b^{2}>0$, the interaction is increased along $\hat{\alpha}=0,\pi$ and reduced in the transverse direction $\hat{\alpha}=\pi/2$. Also, notice that, in the perturbative expansion, the term $\xi b^{2}/6$ corresponds to an isotropic renormalization of the Coulomb coupling, whereas the contribution proportional to $P_{2}(\cos\hat{\alpha})$ produces a quadrupolar angular modulation of the interaction.

\section{Scattering induced by the bumblebee field: timelike configuration}

\subsection{Electron scattering}

The analysis that follows is carried out within the approach introduced in Ref.~\cite{Touati:2024tqy}. In this framework, the Lorentz--violating scattering amplitude $f(\gamma,\xi)$ is obtained by taking as input the interparticle potential derived in Eq.~(\ref{vtimeee}). For the timelike configuration, that potential is
\begin{equation}
V(r)=\frac{1}{4\pi r}\,\frac{1}{1+\frac{\xi b^{2}}{2}},
\label{eq:timelike_potential_scattering}
\end{equation}
which preserves the Coulombic $1/r$ profile and differs from the standard result only by an overall rescaling of the effective coupling. This is the expression obtained in the potential section of the manuscript and should therefore be used consistently throughout the scattering analysis.

Within the first--order Born approximation, the scattering amplitude is written as
\begin{equation}
\label{ampliscat}
f(\gamma,\xi) = -\frac{2m}{\hbar^{2}\gamma}\int_{0}^{\infty} V(r)\,r\sin(\gamma r)\,\mathrm{d}r,
\end{equation}
where $\gamma = 2\kappa \sin\!\left(\frac{\theta}{2}\right)$ and $\kappa$ denotes the wave number of the incident particle. In the present analysis, the interaction potential obtained from the bumblebee propagator is interpreted as the effective potential experienced by a nonrelativistic electron. Accordingly, the parameter $m$ in Eq.~(\ref{ampliscat}) represents the electron mass. Since the Coulomb integral is infrared divergent, we introduce a Yukawa regulator by replacing Eq.~(\ref{eq:timelike_potential_scattering}) with
\begin{equation}
V_{\alpha}(r)=\frac{1}{4\pi r}\,\frac{1}{1+\frac{\xi b^{2}}{2}}\,e^{-\alpha r},
\qquad \alpha>0.
\label{eq:regulated_potential_timelike}
\end{equation}
Substituting this expression into Eq.~(\ref{ampliscat}) and using
\begin{equation}
\int_{0}^{\infty} e^{-\alpha r}\sin(\gamma r)\,\mathrm{d}r
=
\frac{\gamma}{\gamma^{2}+\alpha^{2}},
\end{equation}
we obtain
\begin{equation}
f(\gamma,\xi;\alpha)
=
-\frac{m}{2\pi\hbar^{2}}\,
\frac{1}{1+\frac{\xi b^{2}}{2}}\,
\frac{1}{\gamma^{2}+\alpha^{2}}.
\end{equation}
Taking the limit $\alpha\to 0^{+}$ yields the physical amplitude
\begin{equation}
\label{approVVV}
f(\gamma,\xi)
=
-\frac{m}{2\pi\hbar^{2}\gamma^{2}}\,
\frac{1}{1+\frac{\xi b^{2}}{2}}.
\end{equation}
For small $|\xi b^{2}|$, one may also write
\begin{equation}
f(\gamma,\xi)
\approx
-\frac{m}{2\pi\hbar^{2}\gamma^{2}}
\left(1-\frac{\xi b^{2}}{2}\right).
\end{equation}

\begin{figure}
    \centering
     \includegraphics[scale=0.6]{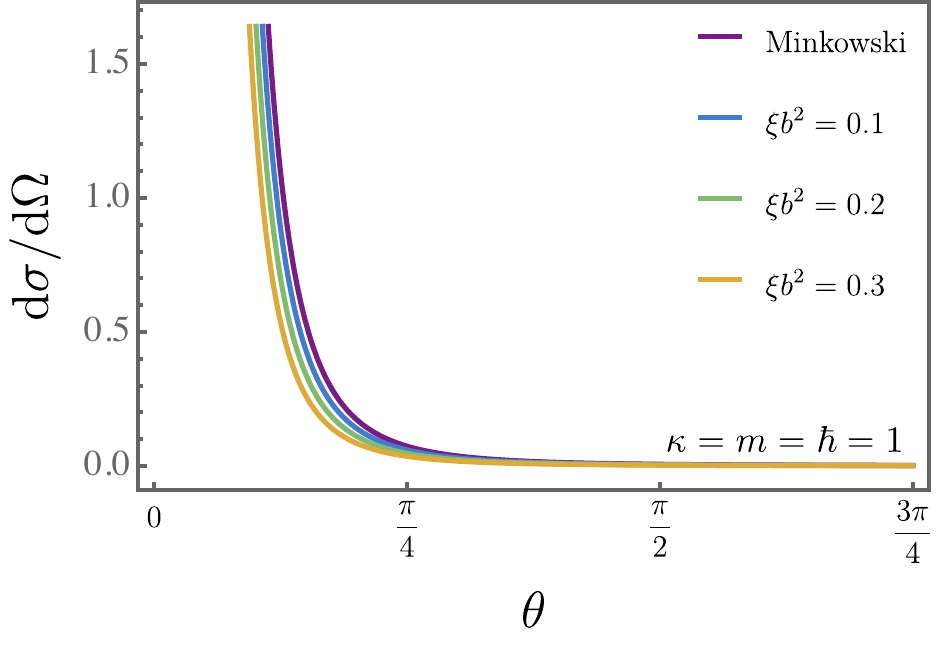}
    \caption{Differential cross section $\mathrm{d}\sigma/\mathrm{d}\Omega$ as a function of the scattering angle $\theta$ for different values of the parameter $\xi b^{2}$.}
    \label{domega}
\end{figure}

The corresponding differential cross section is therefore
\begin{equation}
\label{crosssection}
\frac{\mathrm{d}\sigma}{\mathrm{d}\Omega}
=
|f(\gamma,\xi)|^{2}
=
\frac{m^{2}}{4\pi^{2}\hbar^{4}\gamma^{4}}\,
\frac{1}{\left(1+\frac{\xi b^{2}}{2}\right)^{2}}.
\end{equation}
Using $\gamma=2\kappa\sin(\theta/2)$, this becomes
\begin{equation}
\frac{\mathrm{d}\sigma}{\mathrm{d}\Omega}
=
\frac{m^{2}}{64\pi^{2}\hbar^{4}\kappa^{4}}\,
\frac{1}{\left(1+\frac{\xi b^{2}}{2}\right)^{2}}\,
\csc^{4}\!\left(\frac{\theta}{2}\right).
\label{eq:diff_cross_theta_timelike}
\end{equation}
Thus, the timelike bumblebee background preserves the Rutherford angular dependence, while the Lorentz--violating parameter modifies only the overall magnitude of the scattering. For positive $\xi b^{2}$, the cross section is suppressed relative to the Coulomb case, whereas for negative $\xi b^{2}$ it is enhanced. In Fig.~\ref{domega}, we set $\kappa=m=\hbar=1$.

The total cross section is defined as
\begin{equation}
\sigma=\int \mathrm{d}\Omega\,|f(\theta)|^{2}
=\int_{0}^{\pi}2\pi\sin\theta\,|f(\theta)|^{2}\,\mathrm{d}\theta .
\end{equation}
Taking into account the kinematical relation
$\gamma=2\kappa\sin\!\left(\frac{\theta}{2}\right)$,
we obtain
$\mathrm{d}\gamma=\kappa\cos\!\left(\frac{\theta}{2}\right)\mathrm{d}\theta$
and
\begin{equation}
\mathrm{d}\Omega
=
2\pi\sin\theta\,\mathrm{d}\theta
=
4\pi\sin\!\left(\frac{\theta}{2}\right)\cos\!\left(\frac{\theta}{2}\right)\mathrm{d}\theta
=
\frac{2\pi}{\kappa^{2}}\,\gamma\,\mathrm{d}\gamma .
\end{equation}
Therefore, the total cross section in the first--order Born approximation can be written as
\begin{equation}
\sigma
=
\frac{2\pi}{\kappa^{2}}
\int_{0}^{2\kappa}|f(\gamma,\xi)|^{2}\,\gamma\,\mathrm{d}\gamma .
\end{equation}

Since the amplitude behaves as $f(\gamma,\xi)\propto \gamma^{-2}$, the integrand scales as $\gamma^{-3}$ and the integral diverges at $\gamma\to 0$, corresponding to the forward--scattering limit $\theta\to 0$. This is the standard infrared divergence associated with long--range Coulomb--type interactions. To regularize the expression, we introduce (the cutoff) a minimum momentum transfer
$\gamma_{\min}=2\kappa\sin\!\left(\frac{\theta_{\min}}{2}\right)$,
which physically encodes a finite angular resolution or experimental acceptance. The regulated cross section then reads
\begin{equation}
\sigma(\gamma_{\min},\xi)
=
\frac{2\pi}{\kappa^{2}}
\int_{\gamma_{\min}}^{2\kappa}
|f(\gamma,\xi)|^{2}\,\gamma\,\mathrm{d}\gamma .
\end{equation}
Substituting Eq.~(\ref{approVVV}) yields
\begin{equation}
\sigma(\gamma_{\min},\xi)
=
\frac{m^{2}}{4\pi\hbar^{4}\kappa^{2}}\,
\frac{1}{\left(1+\frac{\xi b^{2}}{2}\right)^{2}}
\left(
\frac{1}{\gamma_{\min}^{2}}
-\frac{1}{4\kappa^{2}}
\right),
\label{eq:sigma_gammamin_timelike}
\end{equation}
or, equivalently,
\begin{equation}
\sigma(\gamma_{\min},\xi)
=
\frac{m^{2}\left(4\kappa^{2}-\gamma_{\min}^{2}\right)}
{16\pi\hbar^{4}\kappa^{4}\gamma_{\min}^{2}}\,
\frac{1}{\left(1+\frac{\xi b^{2}}{2}\right)^{2}} .
\end{equation}

For $\gamma_{\min}\ll\kappa$, the dominant contribution behaves as
\begin{equation}
\sigma(\gamma_{\min},\xi)
\approx
\frac{m^{2}}
{4\pi\hbar^{4}\kappa^{2}\gamma_{\min}^{2}}\,
\frac{1}{\left(1+\frac{\xi b^{2}}{2}\right)^{2}},
\end{equation}
which shows explicitly the strong enhancement due to small--angle scattering. In terms of the angular cutoff, we have
\begin{equation}
\sigma(\theta_{\min},\xi)
=
\frac{m^{2}}
{16\pi\hbar^{4}\kappa^{4}}\,
\frac{1}{\left(1+\frac{\xi b^{2}}{2}\right)^{2}}
\left(
\csc^{2}\!\frac{\theta_{\min}}{2}-1
\right).
\label{eq:sigma_thetamin_timelike}
\end{equation}
These expressions show that the divergence of the total cross section originates entirely from the forward region, while the parameter $\xi b^{2}$ enters only through the overall coupling rescaling.

A convenient dimensionless parametrization of the infrared cutoff is obtained by introducing
$y \equiv \frac{\gamma_{\min}}{2\kappa}\in(0,1]
\,\,\Longleftrightarrow\,\,
\gamma_{\min}=2\kappa y$.
With this definition, the regulated cross section becomes
\begin{align}
\sigma(y,\xi)
&=
\frac{m^{2}}{4\pi\hbar^{4}\kappa^{2}}\,
\frac{1}{\left(1+\frac{\xi b^{2}}{2}\right)^{2}}
\left(
\frac{1}{(2\kappa y)^{2}}-\frac{1}{4\kappa^{2}}
\right)
=
\frac{m^{2}}{16\pi\hbar^{4}\kappa^{4}}\,
\frac{1}{\left(1+\frac{\xi b^{2}}{2}\right)^{2}}\,
\frac{1-y^{2}}{y^{2}}.
\label{eq:sigma_y_def}
\end{align}
In the forward--scattering regime $y\ll 1$, equivalently $\gamma_{\min}\ll\kappa$, this reduces to
\begin{equation}
\sigma(y,\xi)\approx
\frac{m^{2}}{16\pi\hbar^{4}\kappa^{4}}\,
\frac{1}{\left(1+\frac{\xi b^{2}}{2}\right)^{2}}\,
\frac{1}{y^{2}},
\qquad (y\ll 1),
\end{equation}
making the infrared enhancement explicit as a simple $y^{-2}$ divergence. Nevertheless, if we want to express the cutoff through the minimum scattering angle, then using $\gamma_{\min}=2\kappa\sin(\theta_{\min}/2)$ implies $y \equiv \sin\!\frac{\theta_{\min}}{2}\in(0,1]$,
so that Eq.~\eqref{eq:sigma_y_def} is equivalent to Eq.~\eqref{eq:sigma_thetamin_timelike}.

To parametrize the energy dependence, we now define $x \equiv \frac{2\kappa}{m},
\,\,
E=\frac{\hbar^{2}\kappa^{2}}{2m}
=\frac{\hbar^{2}m}{8}\,x^{2}$.
Here it is more appropriate to take simply $x>0$, unless a specific plotting interval is imposed. In terms of $(x,y)$, the cross section can be written as
\begin{equation}
\sigma(x,y,\xi)
=
\frac{1}{\pi\hbar^{4}}\,
\frac{1}{m^{2}x^{4}}\,
\frac{1}{\left(1+\frac{\xi b^{2}}{2}\right)^{2}}
\left(
\frac{1}{y^{2}}-1
\right),
\label{eq:sigma_xy_timelike}
\end{equation}
which separates the infrared behavior, governed by $y\to 0$, from the energy scaling $\sigma\propto x^{-4}$ at fixed $y$.

\begin{figure}
    \centering
     \includegraphics[scale=0.54]{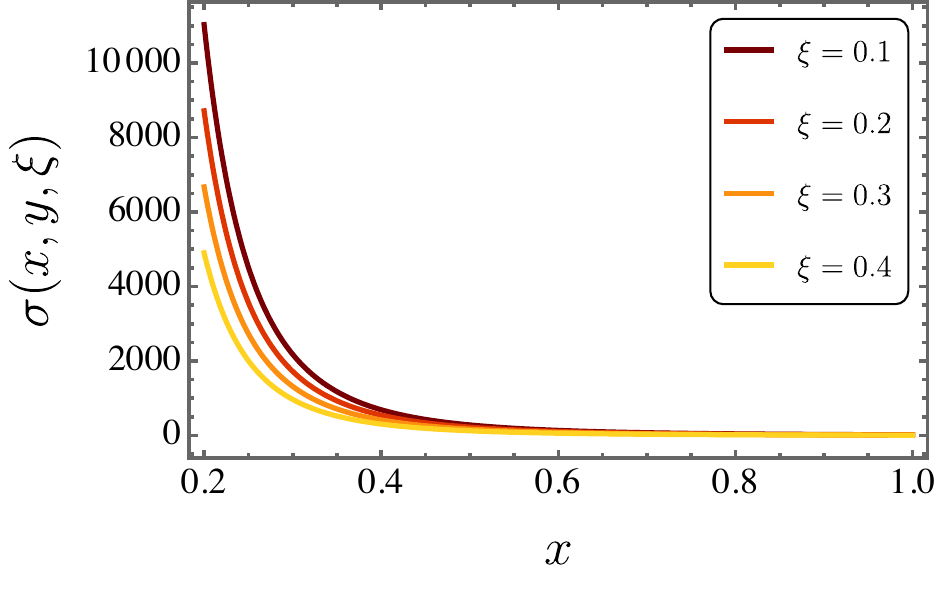}
      \includegraphics[scale=0.53]{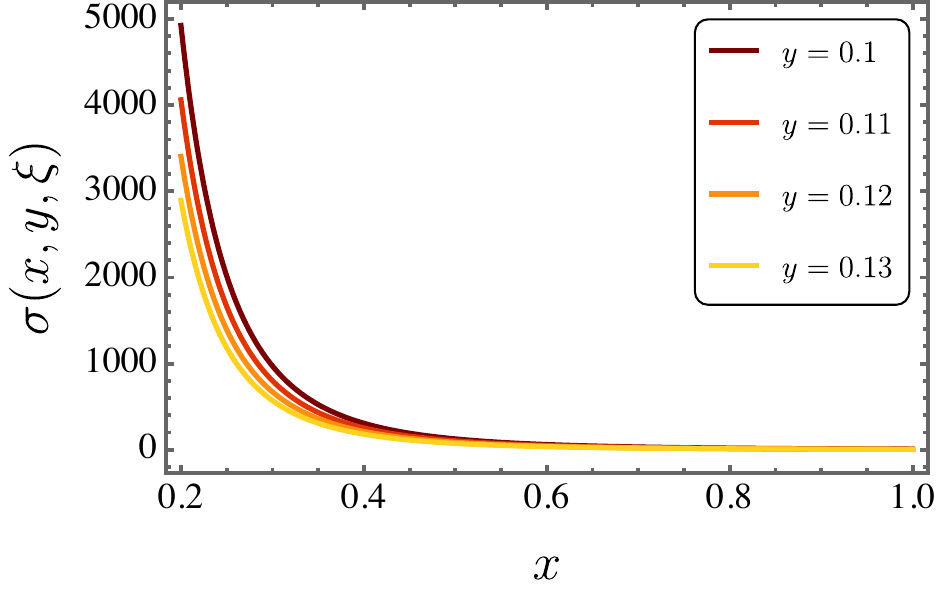}
       \includegraphics[scale=0.5]{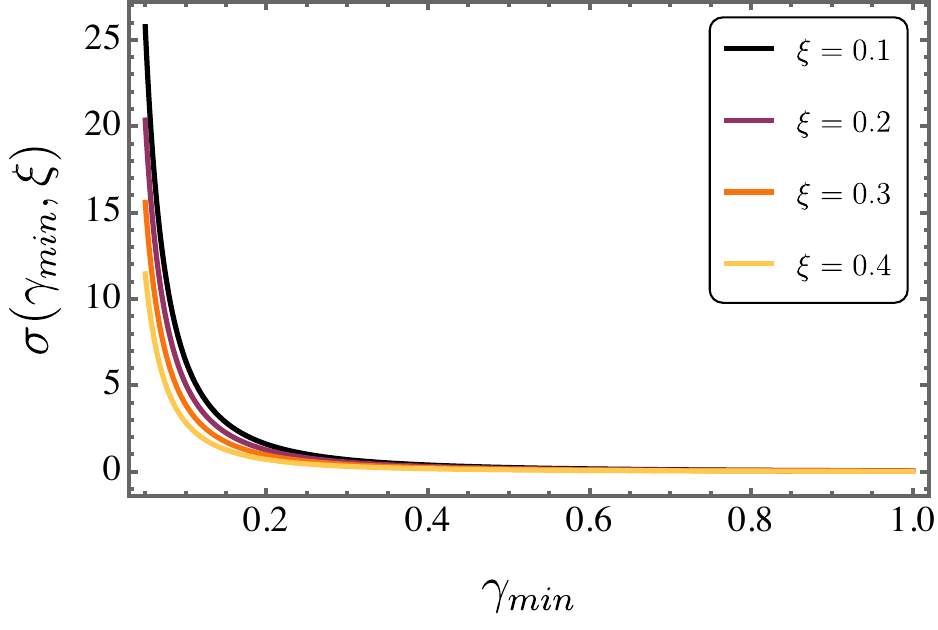}
        \includegraphics[scale=0.5]{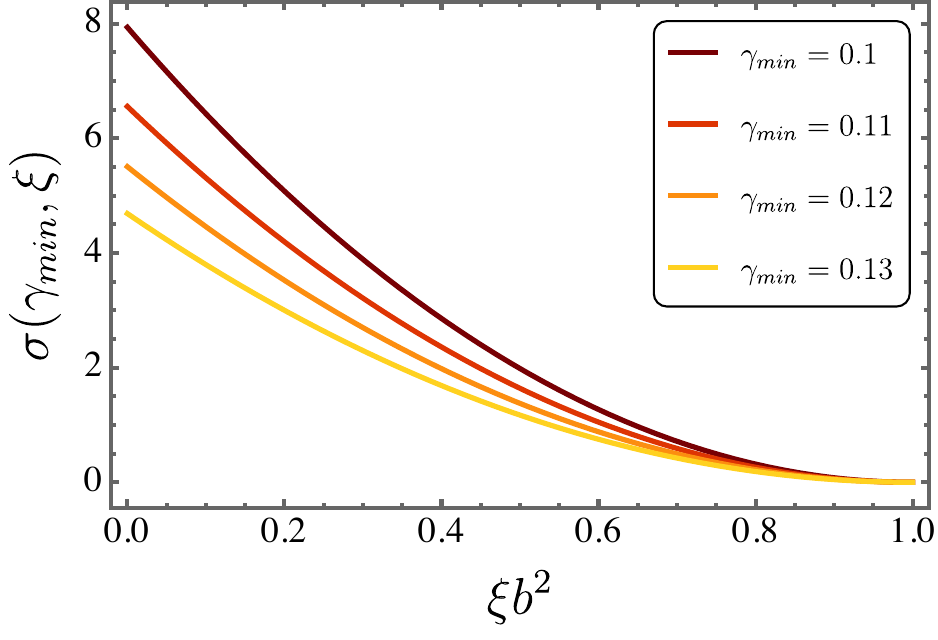}
    \caption{Cross section in different parameter regimes. The upper--left panel shows $\sigma(x,y,\xi)$ as a function of $x$ for several values of $\xi$, with $m=1$ and $y=0.12$. The upper--right panel displays $\sigma(x,y,\xi)$ versus $x$ for different values of $y$, keeping $\xi b^{2}=0.5$ fixed. In the lower--left panel, $\sigma(\gamma_{\min},\xi)$ is plotted as a function of $\gamma_{\min}$ for distinct values of $\xi$, adopting $m=1$ and $\kappa=1$. The lower--right panel presents $\sigma(\gamma_{\min},\xi)$ as a function of $\xi b^{2}$ for several choices of $\gamma_{\min}$, with $m=\kappa=1$.}
    \label{domegamin}
\end{figure}

Fig.~\ref{domegamin} summarizes the behavior of the cross section in different parameter regimes. In the upper--left panel, we display $\sigma(x,y,\xi)$ as a function of $x$ for several values of $\xi$, fixing $m=1$ and $y=0.12$. The upper--right panel shows $\sigma(x,y,\xi)$ as a function of $x$ for different values of $y$, with $\xi b^{2}=0.5$ held constant. In the lower--left panel, we present $\sigma(\gamma_{\min},\xi)$ as a function of $\gamma_{\min}$ for distinct values of $\xi$, adopting $m=1$ and $\kappa=1$. Finally, the lower--right panel illustrates $\sigma(\gamma_{\min},\xi)$ as a function of $\xi b^{2}$ for several choices of $\gamma_{\min}$, again with $m=\kappa=1$. For positive $\xi b^{2}$, the presence of non--metricity suppresses the cross section relative to the Coulomb case, while for negative $\xi b^{2}$ the opposite behavior occurs.

\subsection{High-energy and alternative scattering characterizations}
\label{subsec:highE_eikonal_mott}

The potential in Eq.~\eqref{interpoott} is of Coulomb type, with a bumblebee--induced overall rescaling,
\begin{equation}
V(r)=\frac{1}{4\pi r}\,\frac{1}{1+\frac{\xi b^{2}}{2}}
\equiv \frac{g_{\rm eff}}{r},
\qquad
g_{\rm eff}\equiv \frac{1}{4\pi\left(1+\frac{\xi b^{2}}{2}\right)}.
\label{eq:geff_def}
\end{equation}
As a consequence, the scattering is dominated by the forward region, and any integrated observable requires an infrared regulator, as implemented above.

\subsubsection{Differential cross section in the relativistic regime (Mott-type correction)}
\label{subsubsec:mott}

For relativistic electrons scattered by a static central potential, spin effects introduce a characteristic suppression at large angles. A convenient way to incorporate this correction is to multiply the Rutherford--like result by the standard Mott factor. Keeping the Born normalization and using $\gamma=2\kappa\sin(\theta/2)$, the leading relativistic/spin correction may be encoded as
\begin{equation}
\left(\frac{\mathrm{d}\sigma}{\mathrm{d}\Omega}\right)_{\rm Mott}
=
\left(\frac{\mathrm{d}\sigma}{\mathrm{d}\Omega}\right)_{\rm Born}
\left(1-\tilde{\beta}^{2}\sin^{2}\frac{\theta}{2}\right),
\qquad
\tilde{\beta}\equiv \frac{v}{c}.
\label{eq:Mott_factor}
\end{equation}
Using Eq.~\eqref{eq:diff_cross_theta_timelike}, one finds
\begin{equation}
\left(\frac{\mathrm{d}\sigma}{\mathrm{d}\Omega}\right)_{\rm Mott}
=
\frac{m^{2}}{64\pi^{2}\hbar^{4}\kappa^{4}}\,
\frac{1}{\left(1+\frac{\xi b^{2}}{2}\right)^{2}}\,
\csc^{4}\!\frac{\theta}{2}\,
\left(1-\tilde{\beta}^{2}\sin^{2}\frac{\theta}{2}\right).
\label{eq:Mott_final}
\end{equation}
Therefore, within this approximation, the parameter $\xi b^{2}$ continues to modify only the overall strength, while the angular dependence acquires the usual relativistic/spin modulation through the factor in Eq.~\eqref{eq:Mott_factor}.

\subsubsection{Eikonal formulation (high energy and small-angle scattering)}
\label{subsubsec:eikonal}

An alternative treatment particularly suited to long--range interactions is the eikonal approximation. Introducing the screened potential used in the regularization procedure,
\begin{equation}
V_{\alpha}(r)=\frac{g_{\rm eff}}{r}\,e^{-\alpha r},
\qquad \alpha>0,
\label{eq:Valpha_def}
\end{equation}
the eikonal phase at impact parameter $b$ reads
\begin{equation}
\chi(b)
=
-\frac{1}{\hbar v}\int_{-\infty}^{+\infty}
V_{\alpha}\!\left(\sqrt{b^{2}+z^{2}}\right)\,\mathrm{d}z
=
-\frac{2g_{\rm eff}}{\hbar v}\,K_{0}(\alpha b),
\label{eq:eikonal_phase}
\end{equation}
where $K_{0}$ is the modified Bessel function. The corresponding eikonal amplitude is
\begin{equation}
f_{\rm eik}(\theta)
=
\frac{\kappa}{i}\int_{0}^{\infty}\mathrm{d}b\,b\,
J_{0}(\gamma b)\,
\Big(e^{i\chi(b)}-1\Big),
\qquad
\gamma=2\kappa\sin\frac{\theta}{2},
\label{eq:eikonal_amp_def}
\end{equation}
with $J_{0}$ the Bessel function of the first kind. Expanding $e^{i\chi}\simeq 1+i\chi$ reproduces the leading Born result in the screened theory. Indeed, using
\begin{equation}
\int_{0}^{\infty}\mathrm{d}b\,b\,J_{0}(\gamma b)\,K_{0}(\alpha b)
=
\frac{1}{\gamma^{2}+\alpha^{2}},
\end{equation}
one obtains
\begin{equation}
f_{\rm eik}^{(1)}(\gamma,\xi;\alpha)
=
-\kappa\int_{0}^{\infty}\mathrm{d}b\,b\,J_{0}(\gamma b)\,\chi(b)
= -\frac{2\kappa}{\hbar v}\,g_{\rm eff}\,
\frac{1}{\gamma^{2}+\alpha^{2}}.
\label{eq:eikonal_firstBorn}
\end{equation}
Taking $\alpha\to 0$ yields the expected $1/\gamma^{2}$ behavior,
\begin{equation}
f_{\rm eik}^{(1)}(\gamma,\xi)
=
\lim_{\alpha\to 0}f_{\rm eik}^{(1)}(\gamma,\xi;\alpha)
=
-\frac{2\kappa}{\hbar v}\,g_{\rm eff}\,\frac{1}{\gamma^{2}}
=
-\frac{2\kappa}{\hbar v}\,
\frac{1}{4\pi\left(1+\frac{\xi b^{2}}{2}\right)}\,
\frac{1}{\gamma^{2}}.
\label{eq:eikonal_alpha0}
\end{equation}
This expression is consistent with the behavior of Eq.~\eqref{approVVV}, up to the standard identification between the Schr\"odinger prefactor and the kinematical factor $\kappa/(\hbar v)$ in the high--energy limit.

\subsubsection{Transport (momentum-transfer) cross section}
\label{subsubsec:transport}

For long--range interactions the integrated cross section is dominated by the forward peak and depends strongly on the angular cutoff. A complementary quantity that is often closer to measurable momentum degradation is the transport, or momentum--transfer, cross section,
\begin{equation}
\sigma_{\rm tr}(\theta_{\min},\xi)
\equiv
\int \mathrm{d}\Omega\,(1-\cos\theta)\,\frac{\mathrm{d}\sigma}{\mathrm{d}\Omega}.
\label{eq:sigtr_def}
\end{equation}
Using Eq.~\eqref{eq:diff_cross_theta_timelike} and $1-\cos\theta=2\sin^{2}(\theta/2)$, one finds
\begin{align}
\sigma_{\rm tr}(\theta_{\min},\xi)
&=
2\pi\int_{\theta_{\min}}^{\pi}\mathrm{d}\theta\,\sin\theta\,
\left(2\sin^{2}\frac{\theta}{2}\right)\,
\frac{m^{2}}{64\pi^{2}\hbar^{4}\kappa^{4}}\,
\frac{1}{\left(1+\frac{\xi b^{2}}{2}\right)^{2}}\,
\csc^{4}\!\frac{\theta}{2}
\nonumber\\[4pt]
&=
\frac{m^{2}}{16\pi\hbar^{4}\kappa^{4}}\,
\frac{1}{\left(1+\frac{\xi b^{2}}{2}\right)^{2}}
\int_{\theta_{\min}}^{\pi}\mathrm{d}\theta\,
\sin\theta\,\csc^{2}\!\frac{\theta}{2}.
\label{eq:sigtr_step}
\end{align}
Since $\sin\theta=2\sin(\theta/2)\cos(\theta/2)$, the integral becomes
\begin{equation}
\int_{\theta_{\min}}^{\pi}\mathrm{d}\theta\,
\sin\theta\,\csc^{2}\!\frac{\theta}{2}
=
2\int_{\theta_{\min}}^{\pi}\mathrm{d}\theta\,
\cot\!\frac{\theta}{2}
=
4\left[\ln\!\sin\!\frac{\theta}{2}\right]_{\theta_{\min}}^{\pi}
=
-4\ln\!\sin\!\frac{\theta_{\min}}{2}.
\label{eq:sigtr_int}
\end{equation}
Hence,
\begin{equation}
\sigma_{\rm tr}(\theta_{\min},\xi)
=
\frac{m^{2}}{4\pi\hbar^{4}\kappa^{4}}\,
\frac{1}{\left(1+\frac{\xi b^{2}}{2}\right)^{2}}
\ln\!\left(\csc\!\frac{\theta_{\min}}{2}\right).
\label{eq:sigtr_final}
\end{equation}
Unlike the total cross section, $\sigma_{\rm tr}$ diverges only logarithmically as $\theta_{\min}\to 0$, which makes its infrared sensitivity milder.

\section{Scattering induced by the bumblebee field: spacelike configuration}

\subsection{Electron scattering}
\label{subsec:scattering_spacelike}

For a spacelike background vector $b^\mu$, the scattering analysis must retain the full angular dependence associated with the preferred spatial direction selected by the vacuum expectation value of the bumblebee field. In particular, the anisotropy cannot be incorporated as a constant multiplicative factor, because the relevant angle is determined by the scattering kinematics through the momentum transfer. Moreover, for consistency with the pole structure of the propagator and with the static Green kernel obtained in the potential section, the scattering amplitude must be built from the full spacelike denominator rather than from the simplified on--shell form. 

In the static limit, the momentum--space Green kernel for the spacelike configuration is
\begin{equation}
G(\mathbf q)
=
\frac{1}{q^{2}\left(1-\frac{\xi b^{2}}{2}+\xi b^{2}(\hat{\mathbf q}\!\cdot\!\hat{\mathbf b})^{2}\right)},
\label{eq:Gq_spacelike_scattering}
\end{equation}
where $q=|\mathbf q|$, $\hat{\mathbf q}=\mathbf q/q$, and $\hat{\mathbf b}$ denotes the unit vector along the preferred spatial direction.

Within the first--order Born approximation, the scattering amplitude is obtained from the Fourier transform of the potential,
\begin{equation}
f(\mathbf q,\xi)
=
-\frac{m}{2\pi\hbar^{2}}
\int \mathrm{d}^{3}r\,e^{-i\mathbf q\cdot\mathbf r}V(\mathbf r)
=
-\frac{m}{2\pi\hbar^{2}}\,G(\mathbf q),
\label{eq:Born_amp_general_spacelike}
\end{equation}
which immediately gives
\begin{equation}
f(\mathbf q,\xi)
=
-\frac{m}{2\pi\hbar^{2}}\,
\frac{1}{q^{2}\left(1-\frac{\xi b^{2}}{2}+\xi b^{2}(\hat{\mathbf q}\!\cdot\!\hat{\mathbf b})^{2}\right)}.
\label{eq:fq_exact_spacelike}
\end{equation}

For elastic scattering, $|\mathbf k_{i}|=|\mathbf k_{f}|=\kappa$, and the momentum transfer is $\mathbf q=\mathbf k_{f}-\mathbf k_{i}$ with magnitude
$q=2\kappa\sin\frac{\theta}{2}$,
where $\theta$ is the polar scattering angle.

To display the orientation dependence explicitly, let $\beta$ denote the angle between the incident momentum $\mathbf k_{i}$ and the preferred direction $\hat{\mathbf b}$. Choosing $\hat{\mathbf b}$ along the $z$--axis and taking the incident beam in the $x$--$z$ plane, one finds
$\hat{\mathbf q}\!\cdot\!\hat{\mathbf b}
=
-\cos\beta\,\sin\frac{\theta}{2}
-\sin\beta\,\cos\frac{\theta}{2}\cos\phi$,
where $\phi$ is the azimuthal scattering angle.
Also, it is convenient to define
$\mu(\theta,\phi;\beta)
\equiv
\hat{\mathbf q}\!\cdot\!\hat{\mathbf b}
=
-\cos\beta\,\sin\frac{\theta}{2}
-\sin\beta\,\cos\frac{\theta}{2}\cos\phi$.

Substituting Eq.~\eqref{eq:fq_exact_spacelike} into Eq.~\eqref{eq:Born_amp_general_spacelike}, the scattering amplitude becomes
\begin{equation}
f(\theta,\phi;\beta,\xi)
=
-\frac{m}{8\pi\hbar^{2}\kappa^{2}}\,
\frac{\csc^{2}\!\left(\frac{\theta}{2}\right)}
{1-\frac{\xi b^{2}}{2}+\xi b^{2}\mu^{2}(\theta,\phi;\beta)}.
\label{eq:f_theta_phi_beta_exact}
\end{equation}

Unlike the timelike configuration, the amplitude is not axially symmetric around the beam axis. Besides the usual dependence on $\theta$, it also depends on the azimuthal angle $\phi$ and on the incidence angle $\beta$ that measures the orientation of the incoming beam relative to the preferred direction.

In the perturbative regime $|\xi b^{2}|\ll1$, Eq.~\eqref{eq:f_theta_phi_beta_exact} can be expanded as
\begin{equation}
f(\theta,\phi;\beta,\xi)
=
-\frac{m}{8\pi\hbar^{2}\kappa^{2}}
\csc^{2}\!\left(\frac{\theta}{2}\right)
\left[
1+\frac{\xi b^{2}}{2}-\xi b^{2}\mu^{2}(\theta,\phi;\beta)
\right]
+
\mathcal O\!\left((\xi b^{2})^{2}\right).
\label{eq:f_theta_phi_beta_pert}
\end{equation}

The corresponding differential cross section is therefore
\begin{equation}
\frac{\mathrm{d}\sigma}{\mathrm{d}\Omega}
=
|f(\theta,\phi;\beta,\xi)|^{2}
=
\frac{m^{2}}{64\pi^{2}\hbar^{4}\kappa^{4}}
\csc^{4}\!\left(\frac{\theta}{2}\right)
\frac{1}{\left(1-\frac{\xi b^{2}}{2}+\xi b^{2}\mu^{2}(\theta,\phi;\beta)\right)^{2}}.
\label{eq:dsdo_exact_spacelike_general}
\end{equation}

Expanding again to first order in $\xi b^{2}$ yields
\begin{equation}
\frac{\mathrm{d}\sigma}{\mathrm{d}\Omega}
=
\frac{m^{2}}{64\pi^{2}\hbar^{4}\kappa^{4}}
\csc^{4}\!\left(\frac{\theta}{2}\right)
\left[
1+\xi b^{2}-2\xi b^{2}\mu^{2}(\theta,\phi;\beta)
\right]
+
\mathcal O\!\left((\xi b^{2})^{2}\right).
\label{eq:dsdo_pert_spacelike_general}
\end{equation}

This expression shows that the Rutherford forward enhancement $\propto \csc^{4}(\theta/2)$ is preserved, but the Lorentz--violating correction now depends nontrivially on the full scattering geometry through $\mu(\theta,\phi;\beta)$.

Because the spacelike background breaks axial symmetry, integrated observables must include the azimuthal dependence. A useful quantity is the azimuthally averaged differential cross section,
\begin{equation}
\left\langle \frac{\mathrm{d}\sigma}{\mathrm{d}\Omega}\right\rangle_{\phi}
=
\frac{1}{2\pi}\int_{0}^{2\pi}\mathrm{d}\phi\,
\frac{\mathrm{d}\sigma}{\mathrm{d}\Omega}.
\label{eq:dsdo_phi_avg_def}
\end{equation}

Using
\begin{equation}
\langle \mu^{2}\rangle_{\phi}
=
\cos^{2}\beta\,\sin^{2}\frac{\theta}{2}
+
\frac{1}{2}\sin^{2}\beta\,\cos^{2}\frac{\theta}{2},
\label{eq:mu2_phi_avg}
\end{equation}
one obtains
\begin{equation}
\left\langle \frac{\mathrm{d}\sigma}{\mathrm{d}\Omega}\right\rangle_{\phi}
=
\frac{m^{2}}{64\pi^{2}\hbar^{4}\kappa^{4}}
\left[
\left(1+\xi b^{2}\cos^{2}\beta\right)\csc^{4}\!\left(\frac{\theta}{2}\right)
-
\xi b^{2}\left(3\cos^{2}\beta-1\right)\csc^{2}\!\left(\frac{\theta}{2}\right)
\right]
+
\mathcal O\!\left((\xi b^{2})^{2}\right).
\label{eq:dsdo_phi_avg_final}
\end{equation}

The total cross section remains infrared divergent because of the long--range Coulomb tail. Introducing a minimum scattering angle $\theta_{\min}$, the regulated total cross section is
\begin{equation}
\sigma(\theta_{\min},\beta,\xi)
=
\int_{\theta_{\min}}^{\pi}\mathrm{d}\theta
\int_{0}^{2\pi}\mathrm{d}\phi\,
\sin\theta\,
\frac{\mathrm{d}\sigma}{\mathrm{d}\Omega}.
\label{eq:sigma_reg_def_general_spacelike}
\end{equation}

Using Eq.~\eqref{eq:dsdo_phi_avg_final}, one finds
\begin{align}
\sigma(\theta_{\min},\beta,\xi)
&=
\frac{m^{2}}{16\pi\hbar^{4}\kappa^{4}}
\left(1+\xi b^{2}\cos^{2}\beta\right)
\left(
\csc^{2}\frac{\theta_{\min}}{2}-1
\right)
\nonumber\\[1mm]
&\quad
-
\frac{\xi b^{2}m^{2}}{8\pi\hbar^{4}\kappa^{4}}
\left(3\cos^{2}\beta-1\right)
\ln\!\left(\csc\frac{\theta_{\min}}{2}\right)
+
\mathcal O\!\left((\xi b^{2})^{2}\right).
\label{eq:sigma_reg_final_general_spacelike}
\end{align}

The transport cross section is defined by
\begin{equation}
\sigma_{\rm tr}(\theta_{\min},\beta,\xi)
=
\int_{\theta_{\min}}^{\pi}\mathrm{d}\theta
\int_{0}^{2\pi}\mathrm{d}\phi\,
\sin\theta\,
(1-\cos\theta)\,
\frac{\mathrm{d}\sigma}{\mathrm{d}\Omega}.
\label{eq:sigtr_def_general_spacelike}
\end{equation}

Using again Eq.~\eqref{eq:dsdo_phi_avg_final}, we obtain
\begin{align}
\sigma_{\rm tr}(\theta_{\min},\beta,\xi)
&=
\frac{m^{2}}{4\pi\hbar^{4}\kappa^{4}}
\left(1+\xi b^{2}\cos^{2}\beta\right)
\ln\!\left(\csc\frac{\theta_{\min}}{2}\right)
\nonumber\\[1mm]
&\quad
-
\frac{\xi b^{2}m^{2}}{16\pi\hbar^{4}\kappa^{4}}
\left(3\cos^{2}\beta-1\right)
\left(1+\cos\theta_{\min}\right)
+
\mathcal O\!\left((\xi b^{2})^{2}\right).
\label{eq:sigtr_final_general_spacelike}
\end{align}

For completeness, relativistic spin corrections can be incorporated through the usual Mott factor,
\begin{equation}
\left(\frac{\mathrm{d}\sigma}{\mathrm{d}\Omega}\right)_{\rm Mott}
=
\left(\frac{\mathrm{d}\sigma}{\mathrm{d}\Omega}\right)_{\rm Born}
\left(
1-\tilde{\beta}^{2}\sin^{2}\frac{\theta}{2}
\right),
\label{eq:Mott_general_spacelike}
\end{equation}
where $\tilde{\beta}=v/c$. In other words, in the spacelike case the Mott correction preserves the anisotropic structure of the Born result, while modifying the angular dependence by the usual relativistic spin factor.


\section{Bounds from atomic physics}

\subsection{Timelike configuration of $b^{\mu}$}

\subsubsection{Hydrogen atom data}
\label{sec:bounds_xi_b2_atomic}

In the timelike configuration discussed above, the bumblebee background leaves the
interaction Coulombic and modifies only its overall strength. In our notation,
\begin{equation}
V(r)=\frac{1}{4\pi r}\,\frac{1}{1+\frac{\xi b^{2}}{2}}
\equiv \frac{g_{\rm eff}}{r},
\qquad
g_{\rm eff}\equiv \frac{1}{4\pi\left(1+\frac{\xi b^{2}}{2}\right)},
\label{eq:geff_def_repeat}
\end{equation}
so that, for $|\xi b^{2}|\ll 1$,
\begin{equation}
\frac{\delta g}{g}\equiv \frac{g_{\rm eff}-g_{0}}{g_{0}}
\simeq -\,\frac{\xi b^{2}}{2},
\qquad
g_{0}\equiv \frac{1}{4\pi}.
\label{eq:dg_over_g}
\end{equation}

This impacts hydrogenic bound states because the Coulomb strength fixes both the Bohr
radius and the Rydberg scale. For the attractive potential $V(r)=-g/r$ with $g>0$
(one may take $g\to |g_{\rm eff}|$ in our conventions), the nonrelativistic spectrum is
\begin{equation}
E_{n}=-\frac{\mu\,g^{2}}{2n^{2}},
\label{eq:En_g}
\end{equation}
where $\mu$ is the reduced mass. Therefore, to leading order,
\begin{equation}
\frac{\delta E_{n}}{E_{n}} = 2\frac{\delta g}{g}\simeq -\,\xi b^{2}.
\label{eq:dE_over_E}
\end{equation}

A purely isotropic rescaling of the Coulomb strength can be absorbed into a redefinition
of the electromagnetic coupling used to predict the spectrum. Consequently, a robust
constraint requires combining hydrogen spectroscopy with an independent input for
the coupling strength, equivalently an independent determination of the fine-structure
constant, so as to avoid circularity.

To connect with the atomic-physics literature, it is convenient to parametrize the
Coulomb strength through $\alpha=e^{2}/(4\pi)$, which is simply proportional to $g$.
Let $\alpha_{\rm ext}$ denote an external determination of $\alpha$. In the presence of
Eq.~\eqref{eq:geff_def_repeat}, hydrogen may be viewed as probing an effective coupling
\begin{equation}
\alpha_{\rm H}=\frac{\alpha_{\rm ext}}{1+\frac{\xi b^{2}}{2}}
\simeq \alpha_{\rm ext}\left(1-\frac{\xi b^{2}}{2}\right),
\label{eq:alphaH}
\end{equation}
so that any hydrogen observable whose leading dependence is $\propto \alpha^{p}$ acquires a
fractional shift
\begin{equation}
\frac{\delta \mathcal{O}}{\mathcal{O}} \simeq -\,\frac{p}{2}\,\xi b^{2}.
\label{eq:dO_over_O_general}
\end{equation}
In particular, for gross--structure transition frequencies $\nu\propto \alpha^{2}$,
\begin{equation}
\frac{\delta \nu}{\nu}\simeq -\,\xi b^{2}.
\label{eq:dnu_over_nu}
\end{equation}

Denoting by $\varepsilon_{\rm H}$ the net fractional tolerance in the comparison between a
measured hydrogen frequency and the corresponding theory prediction evaluated with
$\alpha_{\rm ext}$, Eq.~\eqref{eq:dnu_over_nu} yields the indicative constraint
\begin{equation}
|\xi b^{2}| \;\lesssim\; \varepsilon_{\rm H}.
\label{eq:bound_xib2_epsH}
\end{equation}
For an order--of--magnitude estimate, taking $\varepsilon_{\rm H}\simeq \varepsilon_{\alpha}$
set by the external coupling determination, $\varepsilon_{\alpha}\simeq 8.1\times 10^{-11}$,
one finds
\begin{equation}
|\xi b^{2}|\;\lesssim\; 8.1\times 10^{-11}.
\label{eq:bound_xib2_8e11}
\end{equation}
A dedicated analysis should propagate the full hydrogen uncertainty budget and specify
an explicit constant set to avoid double counting.


\subsection{Spacelike configuration of $b^{\mu}$}
\label{sec:bounds_xi_b2_spacelike}

For a purely spacelike background, the static potential becomes anisotropic. Consistently with the potential derived in the previous section, its leading--order form is
\begin{equation}
\nonumber
V(r,\hat{\alpha})
=
\frac{1}{4\pi r}
\left[
1+\frac{\xi b^{2}}{6}
+\frac{\xi b^{2}}{3}P_{2}(\cos\hat{\alpha})
+O\!\big((\xi b^{2})^{2}\big)
\right],
\end{equation}
where $\hat{\alpha}$ is the angle between $\vec r$ and the preferred axis $\vec b$.
Equivalently, using $P_{2}(x)=\tfrac12(3x^{2}-1)$, one may write
\begin{equation}
V(r,\hat{\alpha})
=
\frac{1}{4\pi r}
\left[
1+\frac{\xi b^{2}}{2}\cos^{2}\hat{\alpha}
+O\!\big((\xi b^{2})^{2}\big)
\right].
\end{equation}

For atomic applications we take the attractive Coulomb interaction,
\begin{equation}
V_{\rm H}(r,\hat{\alpha})
=
-\frac{\varrho}{r}
\left[
1+\frac{\xi b^{2}}{6}
+\frac{\xi b^{2}}{3}P_{2}(\cos\hat{\alpha})
\right],
\end{equation}
so that the perturbation relative to $V_{0}(r)=-\varrho/r$ is
\begin{equation}
\label{eq:dV_spacelike}
\delta V(r,\hat{\alpha})
\equiv V_{\rm H}-V_{0}
=
-\frac{\varrho\,\xi b^{2}}{6r}
-\frac{\varrho\,\xi b^{2}}{3r}\,P_{2}(\cos\hat{\alpha})
\;+\;O\!\big((\xi b^{2})^{2}\big).
\end{equation}
This naturally splits into two pieces: an isotropic rescaling proportional to $1/r$ and a
quadrupolar anisotropy proportional to $(1/r)\,P_{2}(\cos\hat{\alpha})$.


\subsubsection{Isotropic piece and $\varrho$--matching}
\label{sec:bounds_xi_b2_spacelike_isotropic}

The first term in Eq.~\eqref{eq:dV_spacelike} is equivalent to an effective coupling
\begin{equation}
\varrho_{\rm eff}
=
\varrho\left(1+\frac{\xi b^{2}}{6}\right),
\qquad
\frac{\delta\varrho}{\varrho}\simeq \frac{\xi b^{2}}{6}.
\end{equation}
Therefore, any hydrogen observable with leading scaling $\mathcal{O}\propto \varrho^{p}$
acquires
\begin{equation}
\label{eq:dO_over_O_spacelike_iso}
\frac{\delta \mathcal{O}}{\mathcal{O}}
\simeq
p\,\frac{\xi b^{2}}{6},
\end{equation}
and in particular, for gross--structure transition frequencies $\nu\propto \varrho^{2}$,
\begin{equation}
\label{eq:dnu_over_nu_spacelike_iso}
\frac{\delta\nu}{\nu}\simeq \frac{1}{3}\,\xi b^{2}.
\end{equation}

As in the timelike case, a pure isotropic rescaling can be absorbed into the definition
of the electromagnetic coupling used in the theoretical prediction. A meaningful bound
therefore requires comparing hydrogen spectroscopy with an independent determination
of $\varrho$ (or an equivalent constant set).

Let $\varepsilon_{\rm H}$ be the net fractional tolerance in the comparison between a
hydrogen frequency and the corresponding theory prediction evaluated with an external
$\varrho_{\rm ext}$. Then Eq.~\eqref{eq:dnu_over_nu_spacelike_iso} gives the indicative bound
\begin{equation}
\label{eq:bound_xib2_spacelike_iso}
|\xi b^{2}|
\;\lesssim\;
3\,\varepsilon_{\rm H}.
\end{equation}
For an order--of--magnitude estimate one may take $\varepsilon_{\rm H}\sim\varepsilon_{\varrho}$,
where $\varepsilon_{\varrho}$ is the relative uncertainty of the external $\alpha$ input.
For instance, atom--recoil interferometry reports $\varepsilon_{\varrho}\simeq 8.1\times 10^{-11}$.
This yields
\begin{equation}
\label{eq:bound_xib2_spacelike_iso_num}
|\xi b^{2}|
\;\lesssim\;
2.4\times 10^{-10},
\end{equation}
with the same caveat as before: a dedicated treatment should propagate the full hydrogen
uncertainty budget and employ an explicit, nonredundant constant set.


\subsubsection{Quadrupolar piece and anisotropy searches}
\label{sec:bounds_xi_b2_spacelike_aniso}

The second term in Eq.~\eqref{eq:dV_spacelike} produces an $m$--dependent shift at first
order in perturbation theory,
\begin{equation}
\label{eq:dE_quad_general}
\delta E^{\rm (quad)}_{n\ell m}
=
-\frac{\varrho\,\xi b^{2}}{3}\,
\Big\langle \frac{1}{r}\Big\rangle_{n\ell}\,
\Big\langle P_{2}(\cos\hat{\alpha})\Big\rangle_{\ell m}.
\end{equation}
For hydrogenic eigenstates $|n\ell m\rangle$, one may use the standard expectation values
\begin{equation}
\label{eq:1_over_r_expect}
\Big\langle \frac{1}{r}\Big\rangle_{n\ell}
=
\frac{1}{a_{0}\,n^{2}}
=
\frac{\mu\,\varrho}{n^{2}},
\qquad
\Big\langle P_{2}(\cos\hat{\alpha})\Big\rangle_{\ell m}
=
\frac{\ell(\ell+1)-3m^{2}}{(2\ell-1)(2\ell+3)},
\end{equation}
valid for $\ell\geq 1$; for $\ell=0$ the quadrupolar expectation vanishes.
Hence, the quadrupolar correction does not affect $S$ states at first order, and it primarily
manifests as:
(i) splittings among magnetic sublevels for $\ell\ge 1$, and/or
(ii) sidereal/orientation modulations if the quantization axis is rotated with respect
to $\vec b$.

A clean observable is the splitting between two $m$ sublevels within a fixed $(n,\ell)$
manifold, for which the isotropic term cancels:
\begin{equation}
\label{eq:splitting_general}
\Delta E^{\rm (quad)}_{n\ell;\,m_1m_2}
=
-\frac{\varrho\,\xi b^{2}}{3}\,
\Big\langle \frac{1}{r}\Big\rangle_{n\ell}\,
\left[
\Big\langle P_{2}\Big\rangle_{\ell m_1}
-
\Big\langle P_{2}\Big\rangle_{\ell m_2}
\right].
\end{equation}
For $\ell=1$ one has $\langle P_2\rangle_{10}=2/5$ and $\langle P_2\rangle_{1\pm 1}=-1/5$,
so the maximal $p$--state splitting is
\begin{equation}
\label{eq:p_state_splitting}
|\Delta E^{\rm (quad)}_{n,\ell=1;\,0,\pm 1}|
=
\frac{\varrho\,|\xi b^{2}|}{3}\,
\frac{1}{a_{0}n^{2}}\,
\frac{3}{5}
=
\frac{\mu\,\varrho^{2}}{5n^{2}}\,|\xi b^{2}|.
\end{equation}
In frequency units,
\begin{equation}
\label{eq:p_state_splitting_freq}
\frac{|\Delta \nu|}{\nu_{n}}
\sim
\frac{|\Delta E|}{|E_{n}|}
=
\frac{2}{5}\,|\xi b^{2}|,
\qquad
E_{n}=-\frac{\mu\varrho^{2}}{2n^{2}}.
\end{equation}

Therefore, any null search for orientation--dependent (or sidereal) modulations of
Zeeman--resolved transition frequencies at fractional level
$\varepsilon_{\rm aniso}$ implies the indicative bound
\begin{equation}
\label{eq:bound_xib2_spacelike_aniso_repeat}
|\xi b^{2}|
\;\lesssim\;
\frac{5}{2}\,\varepsilon_{\rm aniso}.
\end{equation}

Modern clock--comparison and Hughes--Drever type experiments
constrain sidereal variations in atomic transition frequencies
at levels ranging from
\begin{equation}
\varepsilon_{\rm aniso}\sim 10^{-15}
\quad \text{(conservative laboratory bound)}
\end{equation}
to
\begin{equation}
\varepsilon_{\rm aniso}\sim 10^{-18}
\quad \text{(state--of--the--art optical clocks)}.
\end{equation}
Inserting these values into Eq.~\eqref{eq:bound_xib2_spacelike_aniso_repeat}
gives the explicit order--of--magnitude constraints
\begin{equation}
|\xi b^{2}|
\;\lesssim\;
2.5\times10^{-15}
\qquad \text{(conservative)},
\end{equation}
and potentially as strong as
\begin{equation}
|\xi b^{2}|
\;\lesssim\;
2.5\times10^{-18}
\qquad \text{(clock-level sensitivity)}.
\end{equation}

These limits are parametrically stronger than those obtained from the
isotropic rescaling of the Coulomb strength because the quadrupolar term
cannot be absorbed into a redefinition of $\varrho$. Instead, it produces
$m$--dependent splittings and/or sidereal modulations that are directly
observable as violations of rotational invariance in atomic spectroscopy.


\section{Conclusion}

In this work, we investigated \textit{non--metricity} effects on electron scattering within the \textit{metric--affine} formulation of bumblebee gravity, where spontaneous Lorentz symmetry breaking arises from a vector field acquiring a nonzero vacuum expectation value. Treating the affine connection as an independent variable and integrating it out led to an effective Einstein--frame description in which \textit{non--metricity} modified the dispersion relation of the bumblebee perturbations. From the full momentum--space propagator, we identified the pole structure governing the long--range interaction and used it as the basis for the scattering analysis.

Two distinct vacuum configurations were examined. In the purely timelike case, the modified dispersion relation remained isotropic and corresponded to a uniform rescaling of the propagation speed. The resulting interparticle potential preserved its Coulombic $1/r$ behavior and differed from the standard case only through an overall multiplicative factor. Consequently, the scattering amplitude retained the Rutherford angular dependence, while the Lorentz--violating parameter $\xi b^{2}$ entered as a global deformation of the effective coupling. The differential cross section preserved the characteristic forward enhancement, scaling as $1/\gamma^{4}$, whereas the integrated observables displayed the standard infrared sensitivity of long--range interactions. In particular, the total cross section remained dominated by the forward region, while the transport cross section exhibited the expected milder logarithmic dependence on the infrared cutoff. High--energy corrections, including the Mott modification and the eikonal formulation, confirmed that \textit{non--metricity} does not alter the angular structure of the scattering, but only its overall strength.

In contrast, the spacelike configuration selected a preferred spatial direction and rendered the dispersion relation anisotropic. This anisotropy was transferred to the static Green function and led to an interparticle potential that retained its long--range character while acquiring an explicit angular dependence. At leading order in $\xi b^{2}$, the potential displayed an isotropic shift together with a quadrupolar modulation proportional to $P_{2}(\cos\hat{\alpha})$. This structure propagated to the scattering amplitude, producing an orientation--dependent multiplicative factor that modulated the overall magnitude of the cross section while preserving the universal forward enhancement associated with Coulomb--type interactions.
Finally, we discussed phenomenological bounds derived from atomic physics. In the timelike case, hydrogen spectroscopy constrained the isotropic rescaling of the Coulomb interaction, translating experimental precision into bounds on $\xi b^{2}$. In the spacelike configuration, anisotropy searches, such as Hughes--Drever--type experiments and modern clock comparisons, provided stronger complementary constraints by limiting the quadrupolar contribution to transition frequencies. Since this anisotropic term cannot be absorbed into a redefinition of the Coulomb strength, it gives rise to potentially observable $m$--dependent splittings and sidereal modulations.

As a possible extension of the present analysis, it would be worthwhile to examine analogous scenarios in which the graviton obeys the bumblebee--induced modified dispersion relation discussed in Ref.~\cite{Amarilo:2023wpn,Maluf:2014dpa}. Another natural direction is to investigate how the anisotropies associated with the Lorentz--violating background affect thermodynamic functions within the framework of ensemble theory, following the approaches developed in Refs.~\cite{ensemble1,ensemble2,ensemble3,ensemble4,ensemble5}.


\section{Acknowledgments}

\hspace{0.5cm}
A. A. Araújo Filho is supported by Conselho Nacional de Desenvolvimento Cient\'{\i}fico e Tecnol\'{o}gico (CNPq) and Fundação de Apoio à Pesquisa do Estado da Paraíba (FAPESQ), project numbers 150223/2025-0 and 1951/2025. Furthermore, the author is grateful to A.~Yu.~Petrov for valuable and stimulating discussions during the preparation of this manuscript.

\section{Data Availability Statement}

Data Availability Statement: No Data associated in the manuscript


\bibliographystyle{ieeetr}
\bibliography{main}

\end{document}